\documentclass{llncs}
\usepackage{mathtools}
\usepackage{multirow}
\usepackage{graphicx}
\usepackage{subfigure}
\usepackage{url}
\usepackage{hyperref}
\usepackage{csquotes}
\usepackage{todonotes}
\usepackage[ruled,vlined,linesnumbered,slide]{algorithm2e}
\usepackage{placeins}
\usepackage[inline]{enumitem}
\usepackage{listings}
\usepackage{booktabs}
\usepackage{footnote}
\makesavenoteenv{tabular}
\makesavenoteenv{table}
\makesavenoteenv{table*}
\usepackage[ruled,vlined,linesnumbered,slide]{algorithm2e}

\newcommand{\bu}[1]{\textbf{\underline{#1}}}
\newcommand{\ra}[1]{\renewcommand{\arraystretch}{#1}}
\newcommand{\cc}[1]{} 
\newcommand{\mc}[1]{\textit{\color[rgb]{0.4,0.4,0.4}{//#1}}\\}

\newlength\mylen
\newcommand\myinput[1]{%
  \settowidth\mylen{\KwIn{}}%
  \setlength\hangindent{\mylen}%
  \hspace*{\mylen}#1\\}

\begin{document}

\mainmatter  

\title{Hiperfact: In-Memory High Performance Fact Processing - Rethinking the Rete Inference Algorithm}

\author{Conrad Indiono and Stefanie Rinderle-Ma}
\institute{University of Vienna, Faculty of Computer Science, Vienna, Austria\\
\email{\{firstname.lastname\}@univie.ac.at}}

\maketitle

\begin{abstract}

  The Rete forward inference algorithm forms the basis for many rule engines
  deployed today, but it exhibits the following problems: (1) the caching of
  all intermediate join results, (2) the processing of all rules regardless of
  the necessity to do so (stemming from the underlying forward inference
  approach), (3) not defining the join order of rules and its conditions,
  significantly affecting the final run-time performance, and finally (4)
  pointer chasing due to the overall network structure, leading to inefficient
  usage of the CPU caches caused by random access patterns. The Hiperfact
  approach aims to overcome these shortcomings by (1) choosing cache efficient
  data structures on the primary rank 1 fact index storage and intermediate
  join result storage levels, (2) introducing \emph{island fact processing} for
  determining the join order by ensuring minimal intermediate join result
  construction, and (3) introducing \emph{derivation trees} to allow for
  parallel read/write access and lazy rule evaluation. The experimental
  evaluations show that the Hiperfact prototype engine implementing the
  approach achieves significant improvements in respect to both inference and
  query performance. Moreover, the proposed Hiperfact engine is compared to
  existing engines in the context of a comprehensive benchmark.

\end{abstract}

\section{Introduction}
\label{sec:org39ca990}
\label{sec:intro}

The Rete inference algorithm \cite{original_rete} forms the basis for most rule
engines deployed today
\cite{16f19f9427e3d380b0b164833731acfb,4440df1a0c45c9dd3d8ad7c190c6d563,ef4794de8d86f3e0882ea6b473e8d353,7b4bf441c107aa7406e0480be13794de}.
Existing research deal with the extension of the core algorithm for specific
applications
\cite{d521de21f7e948408283be47ca80e55b,c5438a4cbac7b20c278ce34dd3ee1100,2cf321ecbafa720269187402d924b165},
while other approaches cover the deficiencies of the Rete inference approach,
\cite{3b364c343fbb1594558df47b2d56b5c3,06dde6ee870827728c0e31572574d9bb,7a2b382cfad80e438d06946e27075f8d,8ac1b9e0e0ebe78e1c9f23ba31dadaf6}.
The main problems of the Rete inference algorithm can be summarized as
follows: (1) All intermediate results are cached, regardless of the necessity
to do so, leading to ballooning of consumed RAM. (2) All rules passed to the
Rete rule engine will be evaluated, potentially leading to inferred facts that
are irrelevant and might never be used as input, which is the drawback of the
applied forward chaining inference approach. (3) Rete does not deal with join
ordering, which affects the final run-time performance. (4) The network
structure of the Rete algorithm does not facilitate efficient CPU cache usage,
as pointer chasing causes poor usage of prefetched data from the CPU caches
(L1/L2). Regarding (3), join ordering falls under the topic of query
optimization. As efficient fact retrieval is central to identifying rules
having their conditions satisfied, it is critical for inference. Recent work
\cite{91fd3608b3799a815471ec5708af5f6f,b260dee9ff2be53e2e7d030c020d45f9} focus
on efficient query execution, but treats the topic independently from
inference.

The original Rete algorithm \cite{original_rete} proposes the triple fact
structure as the basis for modeling facts, which is the \emph{de facto} data
structure for reasoning within the Semantic Web's Resource Description
Framework (RDF) domain
\cite{f6c08841b199d4e8326ce30f73705bf8,91fd3608b3799a815471ec5708af5f6f}. While
column-oriented storage approaches are discussed in context of RDF / triple
stores \cite{9908f7b1914bdda05e147179e7d2c7c2}, it is usually only within the
context of the fact index layer. The dependent layers on top: intermediate
join result and the final result data structure are not considered. We argue
that by explicitly decoupling the data store (triple store) and the inference
fact processing step, potential performance gains are missed due to not
considering the optimization techniques that are possible within the layers
in-between these two processes. Here both the underlying data storage
techniques and the inference processing technique need to coordinate for
achieving better inference performance. For example in
\cite{f6c08841b199d4e8326ce30f73705bf8}, not all storage proposals support
inference, showing the disconnect of storage design and inference. It is
imperative to consider fact storage techniques in regards to inference as
well. Furthermore, Rete does not consider compression when processing facts as
it assumes that the triples are fully materialized during processing. As shown
in \cite{6291e0a3533703683d5355156812b697}, applying operations directly on
compressed blocks without decompressing has significant impact on the final
performance. Again, all three layers of the storage, i.e. fact index,
intermediate joins, and final result sets, need to coordinate when dealing
with compressed data types.

Thread and data parallelization can both be applied for further improving the
performance of the inference process. Existing works
\cite{456a57403b6029b0f1c5469bac921edd,2cf321ecbafa720269187402d924b165} deal
with parallelizing the core Rete algorithm. In addition to parallelizing the
core fact processing step with multiple worker threads during the fact
processing step, we study where and how data parellelism -- namely
vectorization -- can be applied.

In order to meet the mentioned challenges above we propose Hiperfact, a
comprehensive architecture (data structures and algorithms) for high
performance fact processing covering forward inference and query execution,
allowing monotonic interactive data exploration by supporting incremental
inference. We adapt Stylus' \cite{91fd3608b3799a815471ec5708af5f6f} strong
typing idea both on the fact type level and the data value type level. This
strong typed design allows for improved query execution and parallel fact
insertions. We adopt Inferray's \cite{8f2df16f0f940d37bd677bd56e5ae115} cache
efficiency idea in all levels of the fact storage and employ both thread
parallelism and additionally data parallelism (vectorization), and
column-oriented storage to improve throughput. We do not limit the rule model
to a specific domain and stay as generic as possible. Concretely, the
contributions in this paper tackling the aforementioned problems Rete exhibits
are as follows:

\begin{itemize}
\item \textbf{Fork Join Model Instances and Block Sizing} We illustrate many instances of
the fork join model for thread parallelism and show opportunities for
applying data parallelism in the context of fact processing. These
applications are at the level of both, the single fact and inter-fact
processing layers. Core to the fork join model is determining the data block
size each work thread should take as input.

\item \textbf{Island Fact Processing}. We propose an overall island fact processing method
that, together with sort keys, finds the optimal order for rule processing
and underlying condition lookup and joins. To do so we determine the best
join order for inter-fact processing using cardinality estimates derived
from the underlying rank 1 indices and other condition-associated metrics.

\item \textbf{Derivation Trees}. We define derivation trees in order to detect which
rules can be skipped for processing (laziness property) and identify
parallel write opportunities to the underlying rank 1 indices.

\item \textbf{Evaluation}. We evaluate the proposed techniques and concepts internally
and against other existing rule engine implementations using OpenRuleBench
\cite{dc7be68d49267fd0c3af2b1941810f87}.
\end{itemize}

\subsection{Motivation}
\label{sec:org9b91b0c}

In this section we motivate Hiperfact by summarizing the Rete inference
algorithm and the aforementioned problems.

\begin{figure}[htbp]
\centering
\includegraphics[width=8cm]{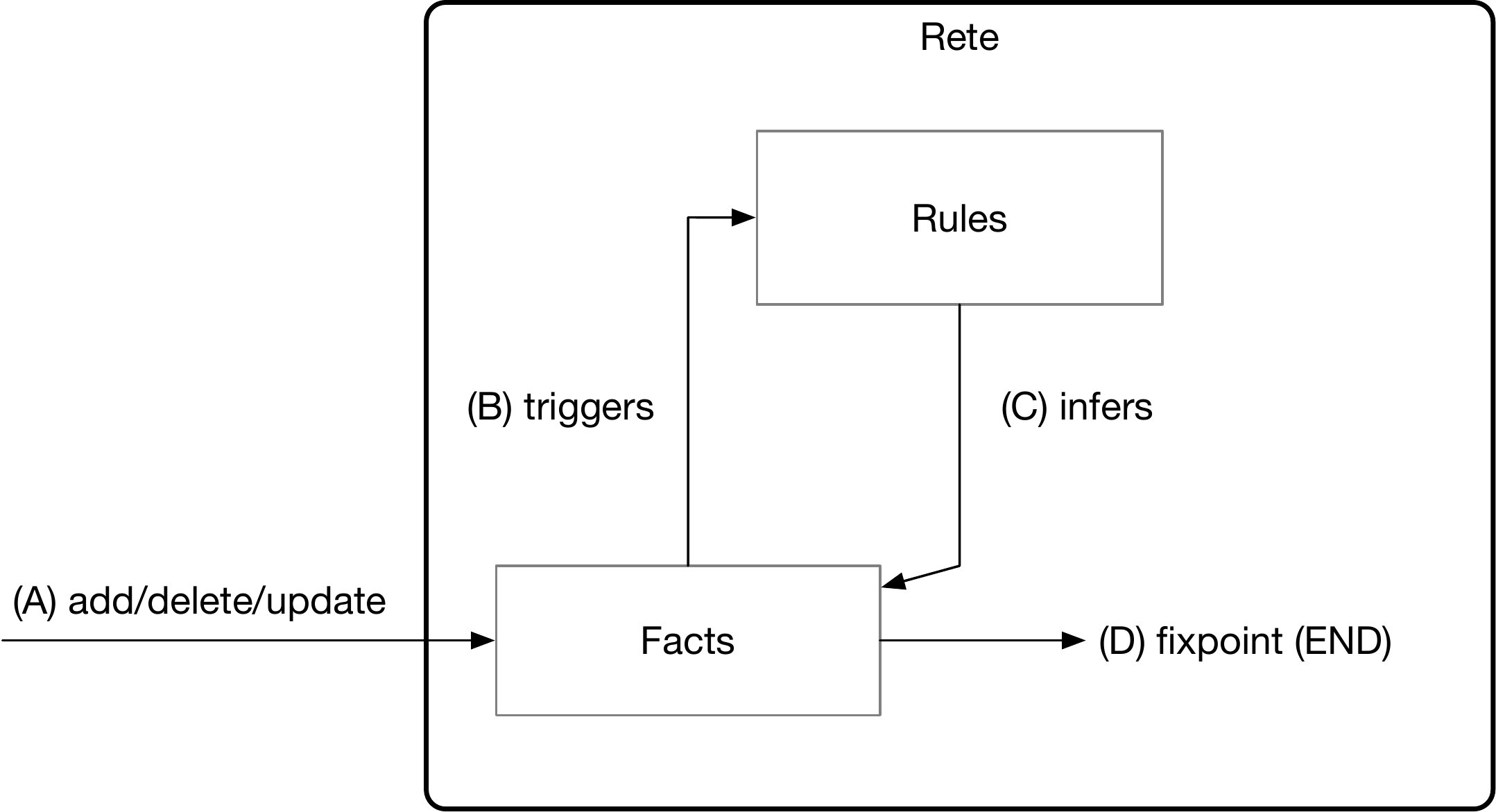}
\caption{\label{rete_loop}Rete Inference Loop}
\end{figure}

The Rete algorithm includes an inference step depicted in Figure
\ref{rete_loop}. The goal of it being to infer derived facts each time facts
are modified (Figure \ref{rete_loop} A). After each modification, user-defined
rules can be triggered (Figure \ref{rete_loop} B). If so, facts can be inferred
(Figure \ref{rete_loop} C), after which the loop continues until a fixpoint is
reached where no more facts can be inferred (Figure \ref{rete_loop} D). At this
point the inference step concludes, waiting until the next fact modification
event occurs. Note that the triggering of rules (B) can be omitted if no rule
conditions are matched, in that case the fixpoint (D) is directly reached.

\begin{figure}[htbp]
\centering
\includegraphics[width=12cm]{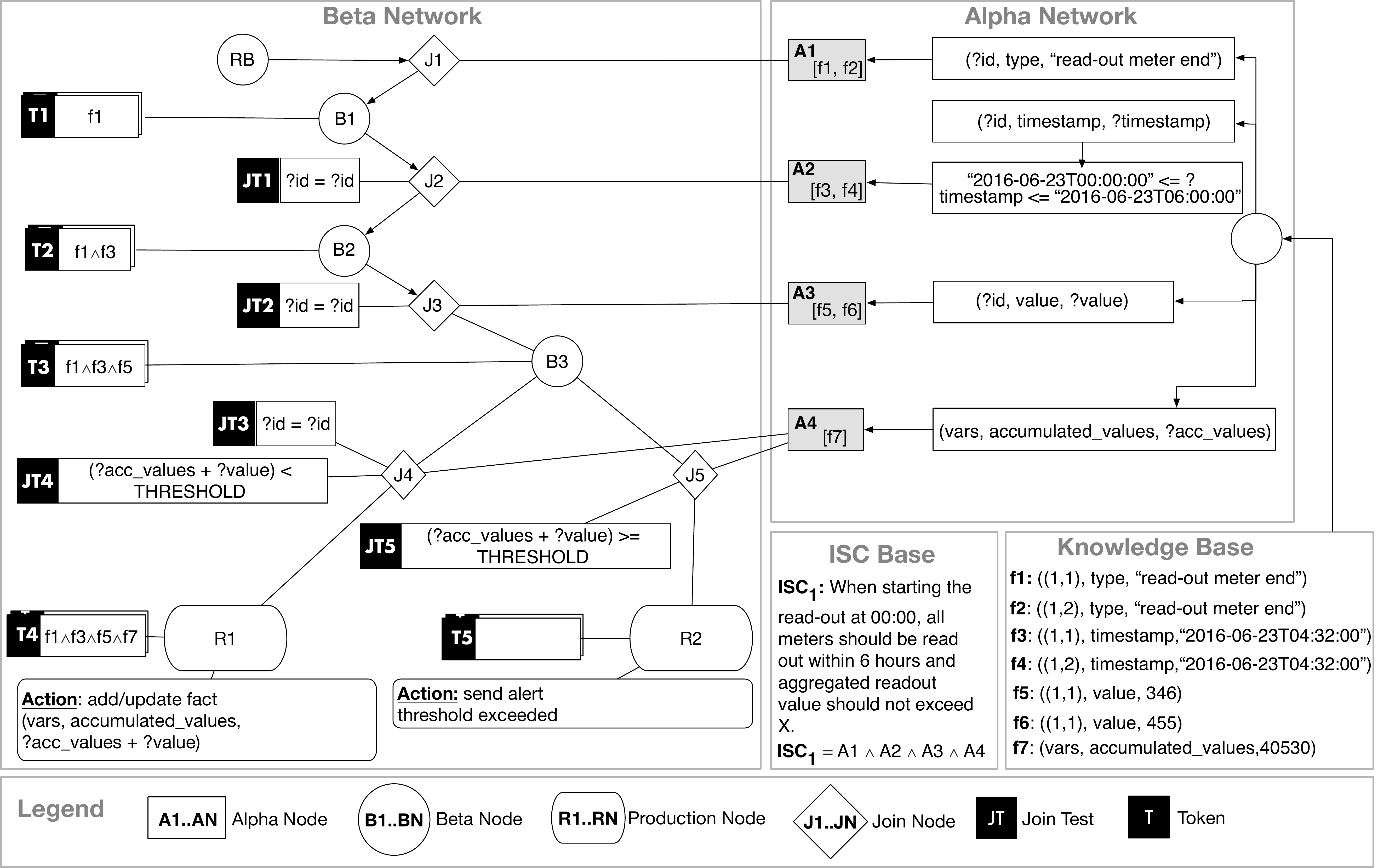}
\caption{\label{rete_basic}Rete Basic Components}
\end{figure}

The data structure of Rete is depicted in Figure \ref{rete_basic} and is
divided into two components: the alpha and the beta network. The alpha
network is designed to handle two issues. First, it is responsible for
indexing the facts entered into Rete, and secondly it breaks down the
user-defined rules into their elementary conditions on which to filter facts
with. This indexing process is conducted based on the triple fact structure:
\texttt{(id, attribute, value)} respected by both the facts being entered as well as
by the conditions making up the rules. This triple structure is divided into
three \texttt{id}, \texttt{attribute}, and \texttt{value} parts. The \texttt{id} marks the unique
identifier to track a certain entity. Each entity has at least one but
unbounded \texttt{attribute}, \texttt{value} pairs. Rule conditions can contain optional
variable parts, prepended with the '?' character. For example, in Figure
\ref{rete_basic} the \texttt{A3} alpha node handles all facts that match the pattern
\texttt{(?id, value, ?value)}, meaning all entities with the attribute literal
\texttt{value} and any kind of associated value on the \texttt{id} and \texttt{value} parts are
passed to the alpha node and processed through the network from there. This
is one of the two pattern matching tasks performed by Rete, in this case the
pattern matching is conducted between fact and conditions, handled by the
alpha nodes.

The beta network handles the second pattern matching task of Rete starting
with facts being passed from the alpha nodes to the join nodes. These join
nodes deal with the joining of facts that match based on common logical
variables. For example, the join node \texttt{J2} joins all facts coming from \texttt{A2}
and \texttt{B1} and joins based on the common variables \texttt{?id}. These variables can
occur in any of the three fact places. As facts are joined together, they are
passed down to the attached beta nodes and stored there as tokens (i.e. \texttt{T2}
in Figure \ref{rete_basic}). Any join nodes attached to these beta nodes are
then triggered, using the newly added tokens as input for the join process.
Triggering those join nodes means that facts are retrieved from the
associated alpha nodes. These facts are then joined with the aforementioned
tokens to perform the join operation. Continuing the previous example, the
beta node \texttt{B2} passes down the tokens to \texttt{J3}, which fetches the facts from
\texttt{A3} (f5 and f6) and performs an equi-join based on the common variables
\texttt{?id}. The results of these join operations are passed to \texttt{B3}. This join
process is repeated until either no more successful joins can be performed or
a production node is activated (see \texttt{R1} and \texttt{R2} in Figure \ref{rete_basic}).
Triggering a production node means the associated rule is executed
(corresponding to Figure \ref{rete_loop} B), after which new facts could be
inferred from (corresponding to Figure \ref{rete_loop} C). Entering these facts
causes the fact processing through the alpha network to be triggered, thus
another fact join processing loop to be started. The fixpoint is then reached
(Figure \ref{rete_loop} D) when the join process in the beta network stops due
to no successful joins being possible. The inference loop of Figure
\ref{rete_loop} is designed to be implemented by both the alpha network and
beta network working together.

\begin{figure}[htbp]
\centering
\includegraphics[width=14cm]{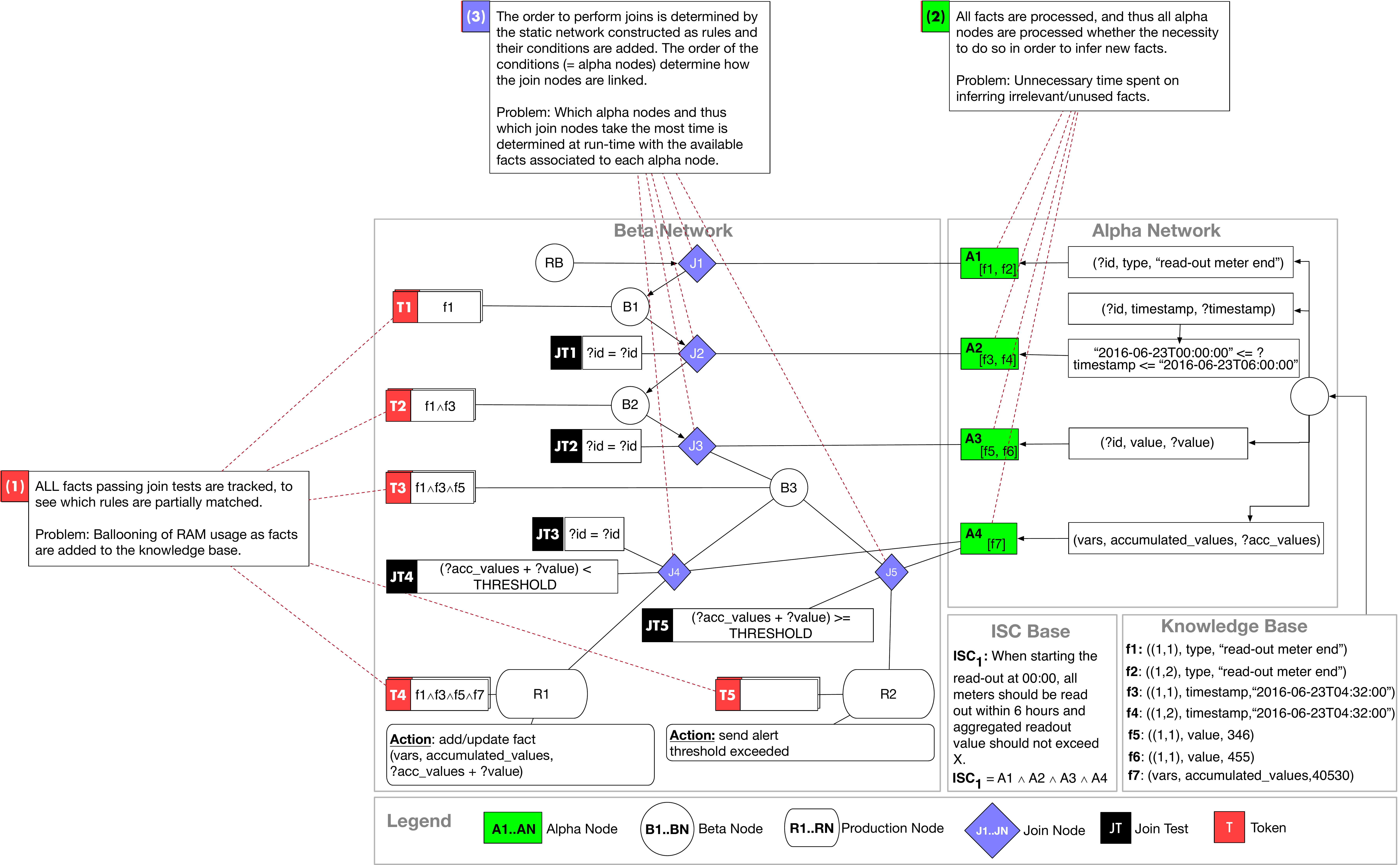}
\caption{\label{rete_problems_annotated}Rete Problems}
\end{figure}

\begin{figure}[htbp]
\centering
\includegraphics[width=12cm]{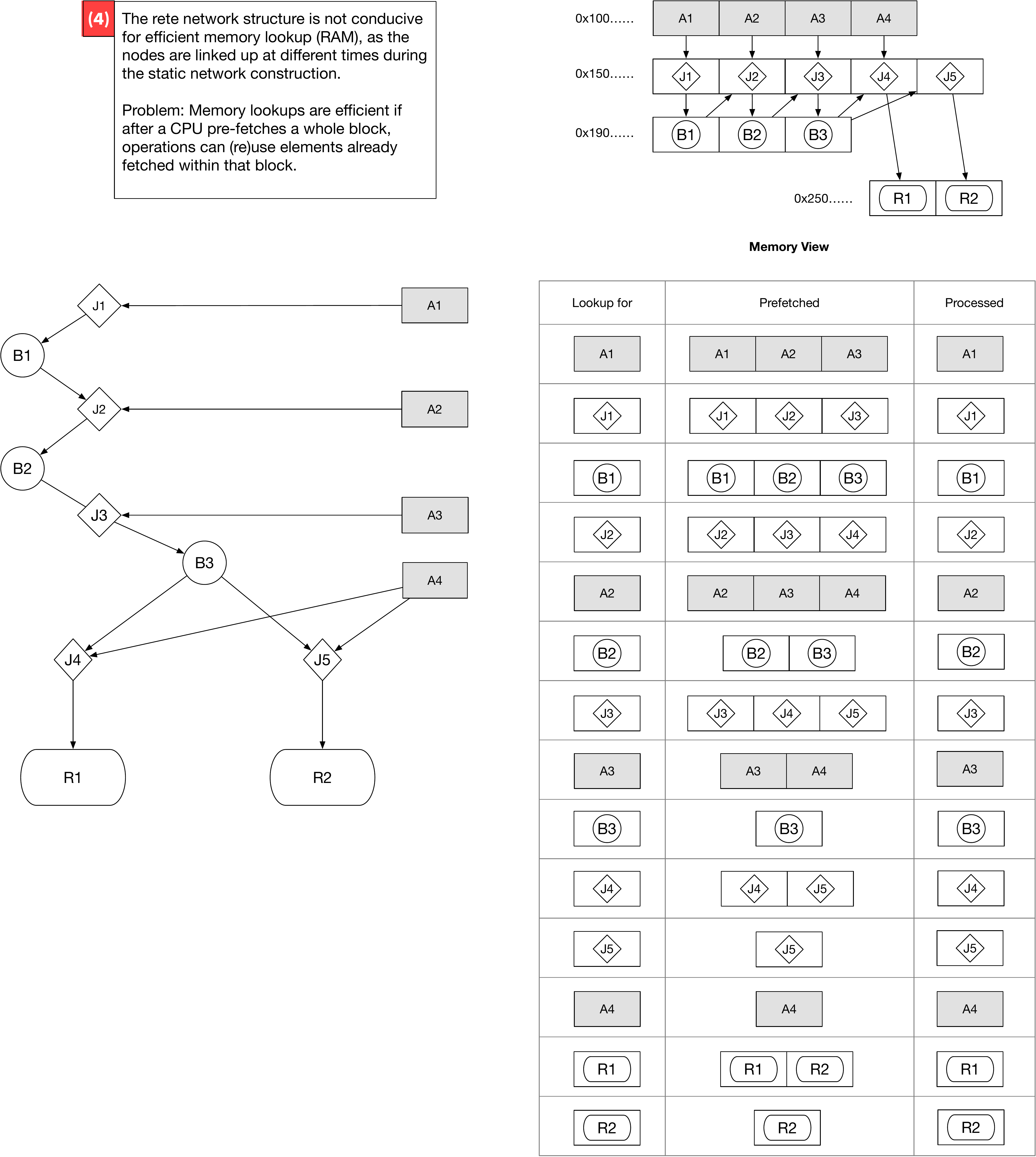}
\caption{\label{rete_memory_access_problems}Rete Memory Access Patterns}
\end{figure}

Based on this foundation we will summarize the common problems Rete has (as
annotated in Figure \ref{rete_problems_annotated}). (\textbf{P1}) The first problem
concerns the tracking of partial rule matches, represented as tokens under
the respective beta nodes. The cost of maintaining these tokens, especially
in a pure in-memory setting, is the ballooning in RAM usage. The design
trade-off being made here is that through sacrificing storage space, the
process of re-joining facts is avoided and thus no wasted joins are
performed. (\textbf{P2}) The second problem is related to the alpha network,
specifically the processing of all facts, regardless of the necessity to do
so in regards to the queries that are in effect. In fact, there is no notion
of queries in Rete. Rete's rule system could be used for defining queries,
where the action part is empty, but this is not explicitly defined. Since
there is no way to connect governing queries with the facts that are required
to answer them, there is no filtering process to know which set of inference
rules are inactive and thus know which facts to ignore. When there is an
ability to skip processing such facts, wasted time spent on those
unneccessary facts can be avoided. (\textbf{P3}) The order to perform joins is
statically pre-determined at the time the network is built at initialization.
This step happens when the rules are added to the Rete system before the
inference loop is triggered. Thus, the order of the rules being added affects
the order of alpha nodes being created. The alpha nodes in turn help build up
the beta network, starting with the definition of join nodes, beta nodes and
finally the production nodes. One critical aspect being ignored in this
regard is the information available at run-time: the actual number of facts
(cardinality) associated to the alpha nodes that actually serve as input in
all of the join operations. Re-ordering the way the facts are joined together
based on this cardinality alone affects the run-time performance of the
inference loop. (\textbf{P4}) The final problem is the incompatibility of Rete's
network structure towards how modern CPUs and RAM work. Figure
\ref{rete_memory_access_problems} shows a full inference step showing the nodes
being processed starting from the alpha node \texttt{A1}. As each node is being
looked up, in modern memory systems \cite{20201204091851} data elements are not
simply looked up atomically, but due to pre-fetching, neighbor entries in the
same array will be fetched as well. The assumption here is that through
spatial locality that those pre-fetched neighbor nodes will be required for
later CPU processing as well. This assumption does not work with Rete, as
only that single node is accessed and processed before another unrelated
lookup (in terms of memory location) is performed for the associated nodes to
continue fact processing (join node \texttt{J1} in this example). The main problem
here is that the pre-fetching behaviour of modern CPUs is not being exploited
in Rete, thus a new structural design is required to exploit that behaviour.
Another related problem is that basic Rete does not prescribe the storage
level design in terms of how the facts are stored. It only prescribes that
the alpha nodes are used to start the processing of facts, but does not
determine how the facts are fetched in which order, nor how the facts are
stored in the knowledge base. These issues are important topics as well, that
we will discuss in this paper. Note that the Rete algorithm is designed to be
an in-memory inference algorithm. Realizing the fact storage in another
context, such as in a persisted or distributed fashion is not part of this
work.

This paper addresses each of the aforementioned problems by redesigning the
core componenets making up the Rete algorithm. The resulting data structures
and algorithms deviate from Rete that we give it a new name under which the
collection of changes constitute a novel inference and querying algorithm:
Hiperfact. In the next section we will describe the baseline Hiperfact
Architecture, which addresses all of the problems mentioned in this section.

\section{Hiperfact Architecture}
\label{sec:org72b09c9}

\begin{figure}[htbp]
\centering
\includegraphics[width=8cm]{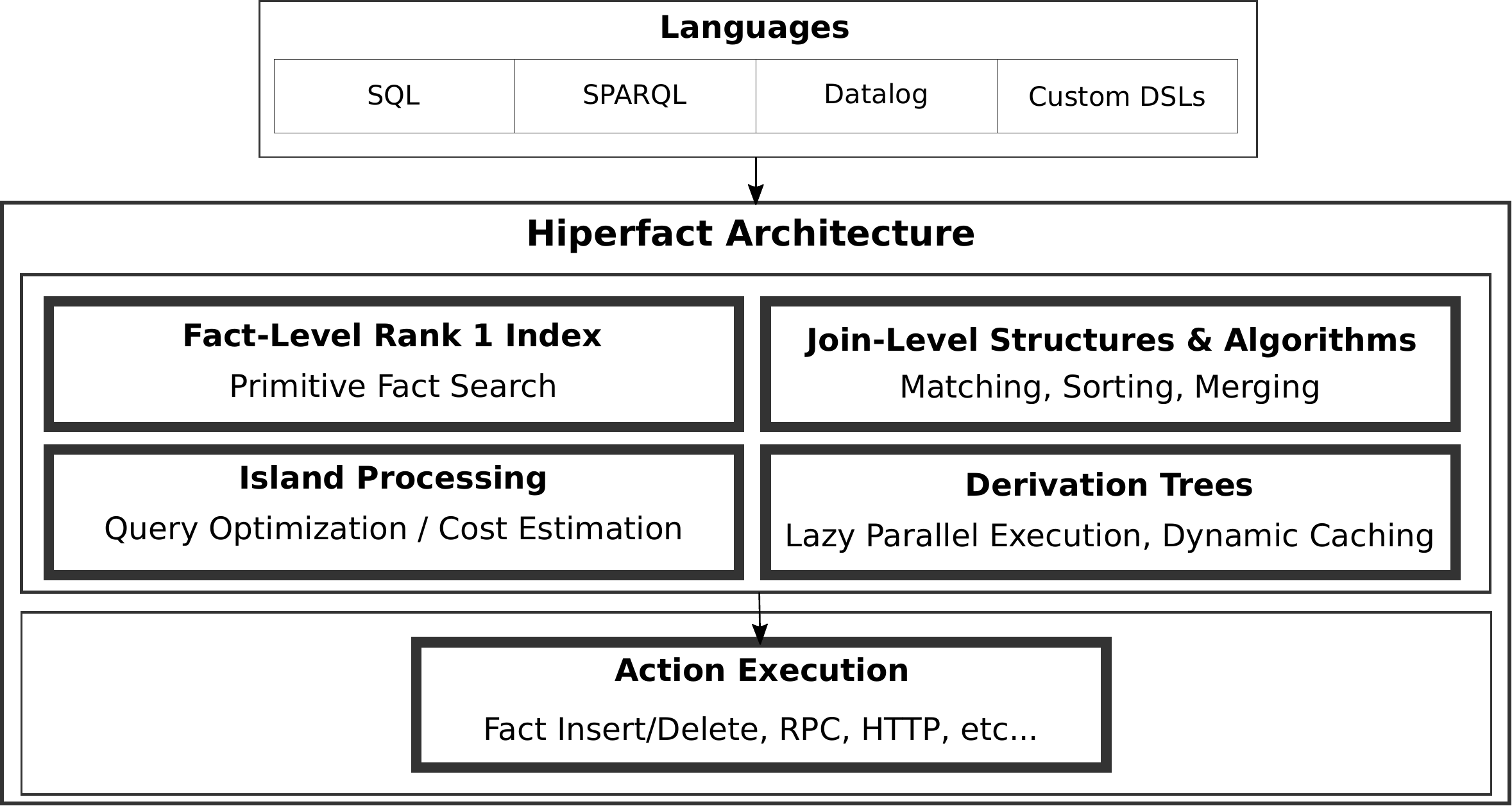}
\caption{\label{hiperfact_architecture_overview}Hiperfact Architecture Overview}
\end{figure}

In this section we highlight the individual components required for realizing
the overall Hiperfact architecture as shown in Figure
\ref{hiperfact_architecture_overview} and discuss relevant technical details. We
follow the design guidelines -- choke points
\cite{36a96c21cb9e42c4bb5ffa0ddd53c243} -- that are critical for achieving good
performance for query execution and thus inference which motivated the chosen
components. Specifically, we focus on (1) estimating cardinality, (2) choosing
the right join order, (3) parallelism and (4) result reuse. The components
discussed in this section will cover some or several of these guidelines.

As shown in Figure \ref{hiperfact_architecture_overview}, the focus of Hiperfact
is less on executing specific languages, such as Datalog or SPARQL, but more
on the underlying low-level process of efficient pattern matching, joining,
processing and aggregating facts before inferring new facts based on matched
results. As long as a translation path to the model of facts, conditions and
rules as defined in this section exists, then the Hiperfact engine can be used
for execution of such queries and rules.

\subsection{Preliminaries}
\label{sec:org07494dd}

We start with preliminary definitions that underly all the components,
starting with the definition of facts.
\begin{definition}{{\small \textbf{[Fact]}}} Based on Rete's \cite{original_rete}
  and RDF's triple fact structure, we extend a fact to be a quintuple
  structured data item to represent a concrete aspect of an event. Facts are
  used as input for Hiperfact to derive more facts, as well as to answer
  queries and has the following structure: $$\texttt{(<fact type> <id>
  <attr> <val> <value type>)}$$ The \texttt{<fact type>} allows facts to be
  strongly typed, which is beneficial for two reasons. First, to store facts
  in distinct namespaces for avoiding unnecessary pattern matches on common
  \texttt{<id>}, \texttt{<attr>} or \texttt{<val>} fact components that are
  not conceptually related. Secondly, it serves as the basis for the
  derivation tree allowing parallel lazy rule evaluation (see Section
  \ref{sec:derivation_tree}). \texttt{<value type>} is one of
  \texttt{\{string, int32, int64, uint32, uint64, float, double, bool\}}.

  \textbf{Example}: Fact (\texttt{City} ``city1'' ``name'' ``New York''
  \texttt{string}) represents the data item with type \texttt{City} holding
  the name of the city ``New York'' under the id ``city1''. Fact
  (\texttt{Person} ``person1'' ``name'' ``Jane'' \texttt{string}) represents
  a person named ``Jane'' stored under the namespace \texttt{Person}.

\label{def:fact}
\end{definition}

\begin{definition}{{\small \textbf{[Condition]}}} A condition specifies a
  pattern to perform fact lookups, returning all facts matching that pattern
  and follows the same structure as a \textbf{fact} with the possibility of
  having named logical variables populating the \texttt{<id>},
  \texttt{<attr>} and \texttt{<val>} fact components. Several conditions can
  be joined together having the same named logical variables. A named
  logical variable starts with the ``?'' symbol. An optional component
  \texttt{<tests>} is added to the condition to house a set of join tests
  (see Definition \ref{def:variable_join_tests}).

  \textbf{Example}: The conditions (\texttt{Person} ?p1 ``livesIn'' ?city
  \texttt{string}) and (\texttt{City} ?city ``name'' ?name \texttt{string})
  perform fact lookups on both \texttt{Person} and \texttt{City} where the
  logical variable ?city match (i.e. are equal). Furthermore, the logical
  variables ?p1 and ?name are mapped to actual person ids and city names
  where those conditions are satisfied.

\label{def:condition}
\end{definition}
\begin{definition}{{\small \textbf{[Rule]}}} A rule holds a collection of
  conditions for specifying the facts the rule
  is to be applied to, and an action part to process the matched facts:
  \vspace{-0.3cm}

  $$(\texttt{conditions = <c1, ..., cN>}, \texttt{action = <a1, ..., aN>})$$

  Actions can be either \texttt{external} for connecting the matched facts
  to external systems (e.g. monitoring/alerting systems) or
  \texttt{internal} for adding new or modifying existing facts. The focus of
  this paper is on internal actions as it pertains to fact processing:
  \texttt{add}(\texttt{new <fact>}), \texttt{delete}(\texttt{old <fact>}),
  and \texttt{replace}(\texttt{old <fact>}, \texttt{new <fact>}).

  \textbf{Example}: Given \texttt{conds} = [(\texttt{DailySales} ?s
  ``profitEUR'' ?p \texttt{double}), (\texttt{DailySales} ?s ``EURUSD'' ?f
  \texttt{double})] and \text{actions} = [\texttt{add}(\texttt{DailySales}
  ?s ``profitUSD'' ($?p\,*\,?f$) \texttt{double})], \texttt{Rule}(conds,
  actions) defines a rule that derives the absolute profit in USD from
  existing facts.

\label{def:rule}
\end{definition}

\subsection{Fact-Level Rank 1 Index}
\label{sec:org70962b5}
\label{sec:rank1_index}

One responsibility of a fact storage layer is the ability to query that
storage layer. In this context, we have a knowledge base that can be queried
to retrieve facts. This is not a topic discussed explicitly by Rete, the only
assumption being that alpha nodes know how to retrieve facts that match the
pattern associated to the alpha node. Especially in an in-memory setting,
that efficient usage of both CPU and modern RAM is critical to keep
performance high, which basic Rete completely ignores. Subsequent fact
processing work depend on this layer to function efficiently. We call this
layer: the \emph{rank 1 index}. The main driver of this component is (\textbf{P4}) where
we try to efficiently use pre-fetched data in order to improve overall fact
lookup operations. The \emph{rank 1 index} is the primary data structure for
storing and retrieving facts. The relevant design choke point \emph{estimating
cardinality} serves as the main guiding principle in designing this
component. In Figure \ref{hiperfact_architecture_overview} we can see that this
component serves as input for the component responsible for query
optimization: \emph{island processing}. For this purpose the \emph{rank 1 index}
provides an API focused on fact retrieval by conditions. As the condition is
the main driver for retrieval of facts, we estimate the cardinality of facts
returned by a given condition thus fulfilling this component's guiding design
principle. First, we need to understand a condition's \emph{rank} that affects the
cardinality of facts to be retrieved.
\begin{definition}{{\small \textbf{[Condition Rank (\texttt{CR})]}}} We define the condition rank (\texttt{CR}) to be equal to the number of
  concrete values filling the \texttt{(<id> <attr> <val>)} triple part of a
  condition. The valid \texttt{CR} range is $[0,3]$, where 0 represents a
  condition that matches all facts. The higher the \texttt{CR}, the more
  specific the filtered facts are. The maximum is 3 representing all triple
  components being filled with concrete values. For example, given the
  condition (\texttt{City} ?id name ?x \texttt{string}), its rank
  \texttt{CR}((\texttt{City} ?id name ?x \texttt{string})) = 1 due to only
  one concrete value found in the attribute part.

\label{def:condition_rank}
\end{definition}

The \emph{rank 1 index} then is a set of inverted indices, one for each triple
part of facts: \emph{id, attribute, value}, allowing for efficient retrieval of
facts queried by conditions that have a \emph{condition rank} of 1 (i.e., have
only one slot filled with concrete values). The corresponding lookup function
used in later components for retrieving such facts is called the \emph{Rank 1
Condition Lookup (R1L)}, defined as follows:
\begin{definition}{{\small \textbf{[Rank 1 Condition Lookup (\texttt{R1L})]}}} Given a condition \texttt{c} with rank $\texttt{CR}(c) = 1$, we define
  three lookup functions \texttt{R1L$_{id}$(c)}, \texttt{R1L$_{attr}$(c)},
  \texttt{R1L$_{val}$(c)}) to be trivial fetch functions from the inverted
  index on the corresponding triple component. The \texttt{R1L} can then be
  defined as:

  \vspace{-0.35cm}

  \[
       \texttt{R1L}(c)= 
         \begin{cases}
             \texttt{R1L$_{id}$}(c),& \text{if } \texttt{CR}(c) = 1 \,\wedge\, !\texttt{isVar}(c.id) \\
             \texttt{R1L$_{attr}$}(c),& \text{if } \texttt{CR}(c) = 1 \,\wedge\, !\texttt{isVar}(c.attr) \\
             \texttt{R1L$_{val}$}(c),& \text{if } \texttt{CR}(c) = 1 \,\wedge\, !\texttt{isVar}(c.val) \\
             \{\},& \texttt{otherwise}
         \end{cases}
  \]

  For example, given \texttt{c} = (\texttt{Person} ?person name ?name
  \texttt{string}) and \texttt{CR}(c) = 1, the lookup \texttt{R1L}(c) fetches
  from the attribute inverted index using the keys (\texttt{Person},
  \texttt{string}, name) and returns the materialized tight array of all
  facts that have \texttt{name} as the concrete value in the attribute triple
  part.

\label{def:rank1_lookup}
\end{definition}

We further define \emph{condition cardinality (CCar)}, to estimate the number of
facts a given condition returns.
\begin{definition}{{\small \textbf{[Condition Cardinality (\texttt{CCar})]}}} The condition cardinality
  is the minimum number of facts indexed on the non-variable triple component
  parts of the condition. Since the condition cardinality is critical for
  determining fact processing order, conditions with \texttt{CR}(c) = 0 need
  to be de-prioritized. Thus, given the set of all triple components $cs = \{
  id, attr, val \}$, the cardinality of a condition is defined:
  \vspace{-0.35cm}

  \[
       \texttt{CCar}(c)= 
         \begin{cases}
             \texttt{inf},& \text{if } \texttt{CR}(c) = 0 \\
             min(\{count(\texttt{R1L$_x$}(c)) \mid \forall x \in cs \}),& \text{otherwise }
         \end{cases}
  \]




\label{def:condition_cardinality}
\end{definition}

What follows are definitions for Condition Lookup functions of \emph{condition
rank} greater than 1 as well as the generic condition lookup function \emph{RL}
allowing for fact lookup independent of \emph{condition rank}.
\begin{definition}{{\small \textbf{[Rank N $>$ 1 Condition Lookup (\texttt{RNL})]}}} Given a
  condition with condition rank $\texttt{CR}(c) > 1$, we define its lookup
  \texttt{RNL}(c) to be an initial \texttt{R1L}(c) lookup on the triple
  component with the lowest cardinality ($count(\texttt{R1L}(c)) =
  \texttt{CCar}(c)$) and performing a subsequent \texttt{EQUAL} filter on the
  other triple components of the returned facts that are not variables. The
  first minimum \texttt{R1L}(c) lookup ensures we start with a small result
  set, as the subsequent filter results are combined in an \texttt{AND}
  relation.

\label{def:rankN_lookup}
\end{definition}
\begin{definition}{{\small \textbf{[Generic Rank Condition Lookup (\texttt{RL})]}}} The generic rank
  condition lookup function (\texttt{RL}) calls the correct rank lookup
  function depending on the condition's rank (\texttt{CR}), and is defined as
  \vspace{-0.35cm}

  \[ \texttt{RL}(c) = \begin{cases} \texttt{R1L}(c), & \text{if } \texttt{CR}(c) = 1\\ \texttt{RNL}(c), & otherwise \end{cases} \]

\label{def:rank_lookup}
\end{definition}

In this section we highlighted the critical functions required for efficient
fact retrieval through \emph{estimating cardinality}, which will be extensively
used in the query optimization phase under the \emph{island processing} component.
For the Hiperfact engine prototype we have three different \emph{rank 1 index}
implementations that expose the common API functions as defined in this
section, focusing on cache-efficient loops when retrieving facts. While we do
try to keep the storage size small for each fact stored, we do not further
optimize on the efficient storage aspect in this work, such as using
compression techniques. In a later section we evaluate these differing index
implementations in regards to their runtime performance measures: loading,
inference, and query time.

The first implementation is a two-level hash table where the first
distinguishes on the \texttt{<fact type>} and the second on the concrete fact
component. Associated to those keys is a tightly packed array of all matching
facts. This can be seen as an inverted index. Since we have three parts to a
single fact, i.e., (\texttt{<id>} \texttt{<attr>} \texttt{<val>}), we define an index for each.
Notable is the \texttt{<id>} index, which can be seen as the primary fact base where
the keys are the \texttt{<fact type>} as well as the unique \texttt{<id>} of the fact.
Additionally, to save more space, the actual fact component for which it is
being indexed can be removed. For example, on the \texttt{<val>} index we eschew the
storage of the \texttt{<val>} part of the quintuple when adding the fact to the
tight array inside the inverted index. The same technique is applied to the
other two indices. At the time of lookup, the full fact quintuple can be
reconstructed adding the missing part using the index key.

The second implementation is a sparse array for holding the two level index.
The main insight stems from the \texttt{<fact type>} exhibiting low randomness and
thus a simple tight array being sufficient for holding the first level index.
The second level index is a sparse array for all possible values being
indexed.

The third implementation follows the same tight array design for the \texttt{<fact
   type>} level and further applies a memory efficient pool of pages that are
pre-allocated and provided dynamically for the second level index. The goal
is to spend less time on dynamic memory operations due to memory
fragmentation, and \texttt{memcpy} operations for accomodating growth at the second
level index.

These three implementations focus on dynamic memory management and efficient
CPU cache usage for fact retrieval. Possible alternative indexing structures
focusing on efficient storage could be bit vectors for further space savings
with the cost of having to perform fact id lookups more frequently when
performing concrete data filtering.

\textbf{String Dictionary}. Strings are a special consideration as they occur
frequently in all facts, i.e., on the \texttt{<id>} and \texttt{<attr>}
level, and in the \texttt{<val>} part in case it is of type \texttt{string}.
The first line of lookups are equality checks, namely when looking up facts
from the inverted index. In order to satisfy that operation and to avoid
dynamic memory allocation performing it, we maintain both a \emph{string to uint64
string-id} radix tree as well as a tightly packed \emph{uint64 string-id to
string} array to index string values. String values are always encoded into
an internal dictionary string-id handle before being added to the inverted
index. This facilitates cache efficient looping through the index at lookup
time as each individual fact becomes fixed size.

\subsection{Island Fact Processing}
\label{sec:org0f87a6b}
\label{sec:island_processing}

Recall that the basic Rete design is a static processing network that does
not take into account the cardinality of inserted facts nor the fact's type.
For this reason, we devised the \emph{island processing} (see Figure
\ref{hiperfact_architecture_overview}) component which focuses on the problems:
\textbf{P1}, \textbf{P3} and \textbf{P4}. On the one hand it eschews the storage vs speed tradeoff
of storing pre-joined tokens, with the reason of having low-cost joins for
taking into consideration \textbf{P4}'s efficient RAM+CPU usage of pre-fetched data
on the underlying \emph{rank 1 index}, as well as the same considerations made in
the design of the \emph{island processing} component. For most use cases,
performing a re-join of existing facts does not hamper performance. In terms
of \textbf{P3}, we build a dynamic processing network based on the underlying \emph{rank
1 index} to process facts in such an order that join effort is kept low.
Several join result data structures are discussed to store intermediate join
results. The driving idea of the \emph{island processing} component is that we can
process facts in sets grouped by their \texttt{id} part (islands of facts that are
directly related) and then joining these islands together. The main problem
becomes in which order to process these islands. This component focuses on
query optimization issues, i.e., determining the most efficient join order
and cost estimation. The latter issue is delegated to the \emph{rank 1 index} and
accumulated into an aggregate cost estimate per island from which the final
join order is derived. Core to an efficient fact processing scheme is the
cost estimation of joins performed for gathering all relevant facts matching
rule conditions. In this section we focus on logical AND relations between
rule conditions, and consider OR relations to be independent paths processed
concurrently.

\textbf{Condition lookup complexity estimation} The basis for the cost estimation
is formed from the lowest rank 1 index level, where the cardinality of
individual conditions are known. For example, given c1 = \texttt{(Book ?x title
    `Title X' string)} and the individual rank 1 cardinalities on the constants
\(count(\texttt{R1L}_{attr}(c)) = 10\) for the attribute `title' and
\(count(\texttt{R1L}_{val}(c)) = 1\) for the value `Title X', we would be
well-advised to start the rank 1 lookup based on \(\texttt{R1L}_{val}(c)\) due
to its higher selectivity. As per Definition \ref{def:condition_cardinality},
this is indeed the operation being chosen, as \(CCar(c)\) equals \(1\) by the
underlying \(\texttt{R1L}_{val}\) lookup. From there the other triple
component constants, or \emph{constant join tests} if they are set, are used for
filtering out facts that match. Notice that \texttt{CCar} does not estimate the
actual final lookup size, but it does help in picking the triple component
with the lowest number of facts stored in the rank 1 index. It is clear that
having a higher rank leads to a lower number of results due to higher amount
of filtering conducted. Thus we can say that we should prefer higher rank
conditions in the beginning of the fact processing to keep the overall join
effort low. Keeping the join result as small as possible at all times
ensures no unnecessary joins are performed at a later stage.
\begin{figure}
\centering
\includegraphics[width=8cm]{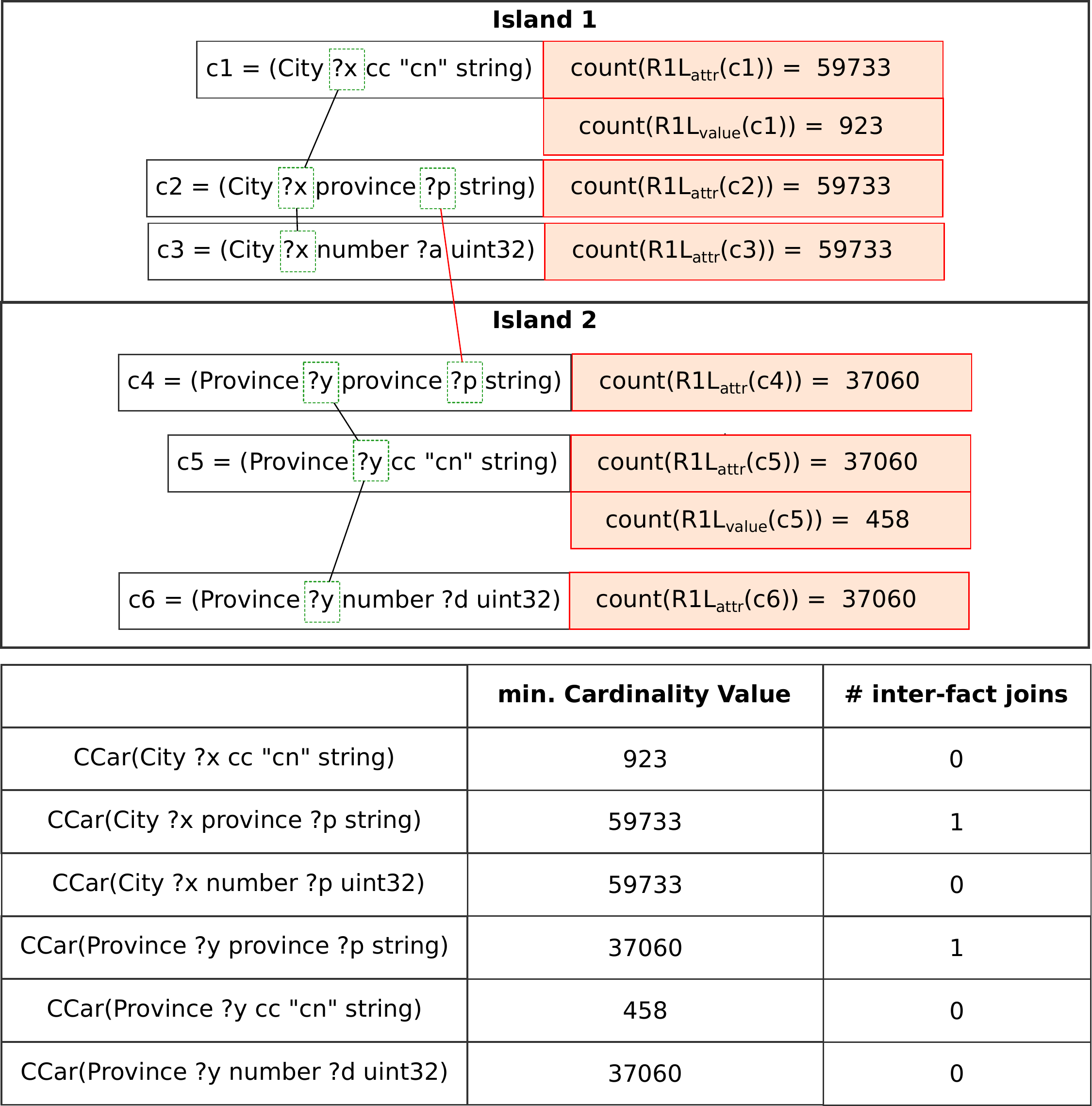}
\caption{Island-based Inference (fact processing)}
\label{island_based_inference}
\vspace{-0.5cm}
\end{figure}

\textbf{Island-based evaluation} After having the cardinality estimate of a single
condition, we can now extend the estimation to a group of conditions. The
conditions inside rules (see Definition \ref{def:rule}) usually come in
distinct island groups. For example, the conditions \texttt{(Book ?x title `Title
    X' string)}, \texttt{(Book ?x description ?d string)}, and \texttt{(Book ?x author `J. D.'
    string)} can be evaluated together, and represent an island, i.e., \textbf{Book}.
\texttt{?x} is bound to all instances of such books. We consider islands to be all
conditions bound to the same variable on the \texttt{<id>} triple component,
loosely linked to other islands through common variables at the \texttt{<value>}
triple component of one island condition and to any other triple position in
one or more other islands. Additionally, for parallel processing purposes,
islands can be evaluated separately. Just as we estimate the cost of a
single condition, we can now estimate the cost of an island by summing up
the individual island conditions. Knowing the island costs before joining
them allows picking the best island to start fact lookups with. Thus, for
the cost estimation of this particular example, the aggregation would be
defined as

\begin{equation} \label{formula:island_cardinality}
\begin{split}
  island_{1} &= \texttt{CCar((Book ?x title `Title X' string))} \\
             &+ \texttt{CCar((Book ?x description ?d string))} \\
             &+ \texttt{CCar((Book ?x author `J. D.' string))}
\end{split}
\end{equation}

\textbf{Example}. Figure \ref{island_based_inference} shows two islands bound to the
\texttt{<id>} variables \texttt{?x} and \texttt{?y}. The rank 1 and subsequent rank 2 cardinality
cost estimations are shown as well. The aggregate island costs for \texttt{?x} is
higher than \texttt{?y}'s. For that reason, \texttt{?y} is evaluated first. The condition
lookup sequence for \texttt{?y} becomes: \texttt{(Province ?y cc `cn' string)} [cost:
458], \texttt{(Province ?y province ?p string)} [cost: 37060], \texttt{(Province ?y number
    ?d uint32)} [cost: 37060]. We additionally track the number of \emph{inter-fact
joins} between conditions. This number helps us to identify the hook point
on which to perform joins between islands. Having evaluated island \texttt{?y} we
can start with island \texttt{?x}. Then, we continue with the condition that links
to island \texttt{?x}. This would be the condition with the only number of
\emph{inter-fact joins} \(> 0\), in this case: \texttt{(Province ?x province ?p string)},
which includes the unbound variable \texttt{?p} from the previous island
evaluation. Thus the full condition lookup sequence for island \texttt{?x} becomes:
\texttt{(City ?x province ?p string)} [cost: 59733], \texttt{(City ?x cc `cn' string)}
[cost: 923], \texttt{(City ?x number ?a uint32)} [cost: 59733]. Note that the first
condition's rank increases at lookup time as the variable \texttt{?p} is bound and
the values are now known. Instead of performing a lookup with cost 59733, we
can perform the \texttt{RNL} lookup of rank 2: \texttt{RNL}(\texttt{(City ?x
    province bound\_p string)}) \(\texttt{count}(?p)\) times, and then combining
the results.

\begin{figure}[htbp]
\centering
\includegraphics[height=4cm]{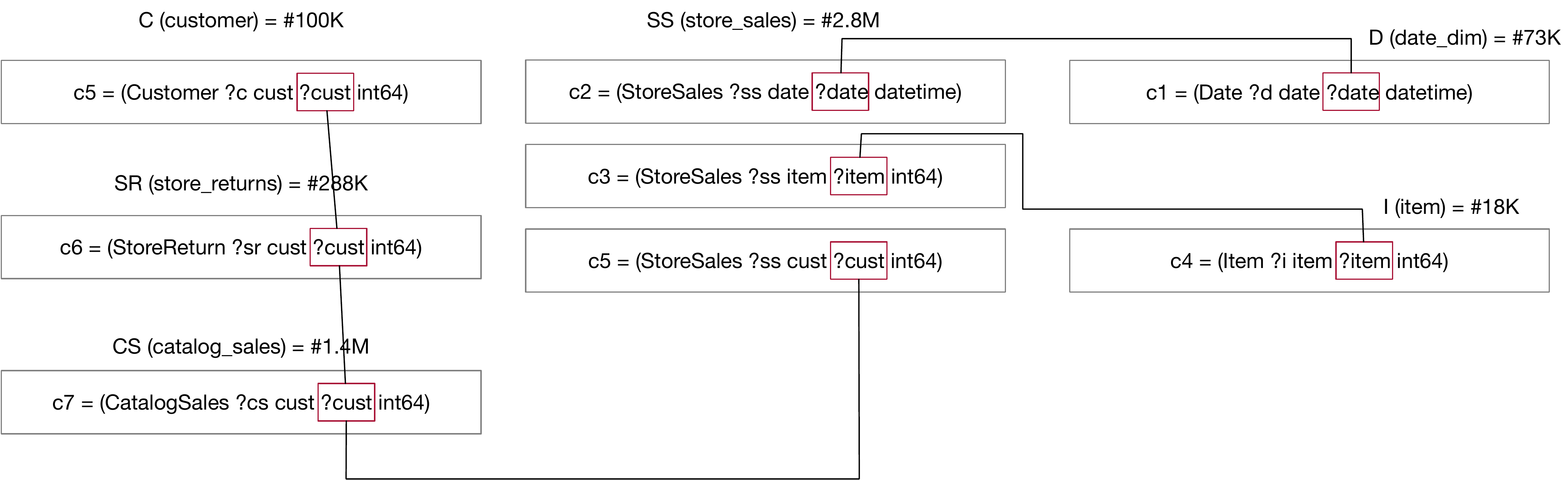}
\caption{\label{tpc_island_based_inference}TPC Island-Processing Example}
\end{figure}

\textbf{Example: TPC}. Figure \ref{tpc_island_based_inference} shows a real-world
benchmark example (TPC) for OLAP-based systems. In this particular use case,
a shopping data set with \texttt{customers}, \texttt{sales}, \texttt{store\_sales} are modeled.
Applying the island processing approach to this data set for a query dealing
with customer returns would identify the following islands: \texttt{C} with
cardinality=100K, \texttt{SR} with cardinality=288K, \texttt{CS} with cardinality=1.4M,
\texttt{SS} with cardinality=2.8M, \texttt{D} with cardinality=73K, \texttt{I} with
cardinality=18K. The first three islands (C, SR, CS) would be processed
first in order of increasing cardinality, and then the final main island
(SS). Even though the last two (D,I) have the lowest cardinality, they are
only required for the SS island, and are thus delegated until that island is
processed.

We can now define the island-based evaluation algorithm (see Algorithm
\ref{algorithm:island_based_evaluation}), which is split into six phases.
\textbf{Phase 1} is concerned with collecting all conditions that occur in the rule
and mapping them according to the \texttt{<id>} variable, associating to it the
following metrics: condition rank, number of inter-fact joins and rank 1
cardinality values per \texttt{<id>}, \texttt{<attr>}, \texttt{<val>} triple component (see
Definition \ref{def:condition_cardinality}). Conditions are strongly typed,
meaning at rule definition time the rule writer has to explicitly set the
\texttt{<fact type>}. This reduces ambiguous condition lookups where the same
attribute occurs in many namespaces. In \textbf{Phase 2}, all the mapped \texttt{<id>
    variables} are iterated and for each associated set of conditions an island
is initialized, associating to it the total estimated cost (see Equation
\ref{formula:island_cardinality}), as well as outgoing and incoming variables.
The variables are required in order to know how to join other islands to
this one. Additionally, we need to track which condition is mapped to this
island. \textbf{Phase 3} is the main join phase. It starts by sorting the islands
by their total cost, and then starting lookups for the first island. When
processing an island, conditions are sorted by their cardinality. The rank
lookup function (\texttt{RL}) refers to the one defined in Definition
\ref{def:rank_lookup}, performing rank 1-3 lookups depending on the
condition's rank. The \emph{join} function refers to the join method as defined
in Section \ref{sec:join_algorithm_instance}, and includes the logic to choose
the appropriate join type. The core part is to join all the separate
conditions belonging to the same island on the \textbf{id} position so that the
final island can be materialized at a later point. If there is a next island
to be joined, its incoming variables are inspected, matched with the
corresponding variable name that can be joined. The actual condition and the
triple position where to join with is marked and prepared as well (via
sorting, such that the corresponding condition is in front). Phase 3 ends
once all islands are processed. \textbf{Phase 4} then materializes and builds the
final data structure, allowing user defined actions to be triggered and
processed (see "Action Execution" component in Figure
\ref{hiperfact_architecture_overview}). In the original Rete algorithm, all
pattern matches for each rule are performed first and rules activated in
that process collected in a so called \emph{conflict set} before the associated
actions (e.g., inferring new facts) are triggered. The order in which these
rule actions are executed is based on the rule writer's priority
designation, usually a positive integer number where the lower number
represents higher priority. In Hiperfact, we follow the derivation trees'
(see Section \ref{sec:derivation_tree}) order of evaluation where the
dependency of inferred facts to associated rules are explicitly modeled,
allowing for concurrent write access to the rank 1 indices. Thus all
inferred facts and any actions associated to the current rule are executed
once phase 5 is reached. In \textbf{Phase 5} cleanup operations are performed,
mainly releasing stale memory, especially deallocating the final result
structure, which has been processed at this point.
\begin{algorithm} [!h]
\begin{scriptsize}
\KwIn{}
\myinput{$rule$ - \emph{Single Rule to be evaluated}}
\KwSty{Begin}\\

  \mc{ Phase 1: Hash and aggregate statistics on conditions to determine
  islands and best join order. }
  id\_hash $\leftarrow$ \{\};
  attr\_hash $\leftarrow$ \{\};
  value\_hash $\leftarrow$ \{\};
  island\_tracker $\leftarrow$ \{\};
  \\
  \ForEach{ $condition$ in $rule.conditions$ }
  {
    cond $\leftarrow$ initialize condition: hash, rank, connected level, rank 1 cardinality values per triple component;
    \\
    $\{ id\_hash, attr\_hash, value\_hash \} \leftarrow$ hash cond varname to its triple position;
  }

  \mc{ Phase 2: Detect all islands and sort by total estimated cost. }
  $islands \leftarrow \{\};$\\
  \ForEach{ $varname$ in $id\_hash.keys()$ }
  {
    conds $\leftarrow id\_hash[varname]$ 
    \\
    island $\leftarrow$ initialize island from conds: total estimated cost, register outgoing vars, ingoing vars;
    \\
    $islands.add(island)$

    \ForEach{ $cond$ in $conds$ }
    {
      it $\leftarrow$ island\_tracker[cond.hash] $\mid$ initialize island tracker from cond: count = 0; \\
      island\_tracker[cond.hash] $\leftarrow$ it.count + 1;
    }
  }

  \mc{ Phase 3: Process each island, connected through common variables.  }
  sort(islands by total\_cost); \\
  sort(islands[0].conditions by cardinality, connected level); \\
  join\_result $\leftarrow$ prepare\_join(\texttt{RL}(islands[0].conditions[0]));\\
  \ForEach{ $island$ in $islands[1:]$ }
  {
    \ForEach{ $cond$ in $island.conditions[1:]$ }
    {
      rhs $\leftarrow$ \texttt{RL}(cond);\\
      join\_result $\leftarrow$ join(join\_result, rhs, POS\_ID);\\
      \texttt{hash\_combine}(join\_result, cond);\\
    }
    \If{ has\_next\_island(island) }
    {
      connected\_vars $\leftarrow$ determine whether cond vars exist for joining to next island using out\_vars; \\
      \If { length(connected\_vars) $>$ 0 }
      {
        sort(connected\_vars by cardinality, connected level);\\
        join\_pos $\leftarrow$ connected\_vars[0].join\_pos;
      }
      \Else
      {
        join\_pos $\leftarrow$ POS\_ID;
      }

      sort(next\_island.conditions by cardinality, connected level);\\

      rhs $\leftarrow$ \texttt{RL}(next\_island.conditions[0]); \\
      join\_result $\leftarrow$ join(join\_result, rhs, join\_pos);\\
      \texttt{hash\_combine}(join\_result, cond);\\
      island $\leftarrow$ next\_island;
    }
  }

  \mc{ Phase 4: Trigger associated rule actions }
  result $\leftarrow$ build\_join\_results(join\_result); \\
  rule.action(result);

  \mc{ Phase 5: Cleanup.  }
  clean(result);

\caption{Island-based Evaluation Algorithm}
\label{algorithm:island_based_evaluation}
\end{scriptsize}
\end{algorithm}

\subsubsection{Sort Keys}
\label{sec:org6981dd6}
\label{sec:sort_keys}

The previous fact processing approach assumes the following: (a) that the
metrics used for ordering the fact evaluation are fixed, and (b) that the
sort order for that particular set of metrics is pre-determined. Namely,
these are the condition rank, connected level of variables, condition triple
component cardinality values, and estimated island cost. With \emph{sort keys} we
extract the ordering logic from the island fact processing algorithm,
allowing us to change the set of metrics and the sort order independent of
the fact processing algorithm. This decoupling allows us to observe the
effect of metric sorting on the overall performance of the inference
algorithm, and reduce the number of sort calls made in general (see Section
\ref{sec:derivation_tree}: \emph{derivation trees}). For sorting the sort keys we
can apply the parallel sort merge algorithm introduced in Section
\ref{sec:fork_join_sort_merge_instance}, as a \emph{sort key} is defined as a
32-bit unsigned integer value that is populated with the different metrics
in specific bit ranges of that value.

\textbf{Capping sort key buckets.} Before defining the bit ranges, we first
discuss the need to bucketize and lower the entropy of the \emph{sort key}.
Instead of using the actual values for filling the individual sort key
buckets, which potentially might increase the available bucket size we can
opt to keep the relative order between the actual values but mapping them
to buckets. For example, having the distinct estimated island scores \(\{
     2043, 6833, 6833, 9700, 50900, 160000, 700000 \}\) we observe that fitting
the maximum value of 70000 requires 20 bits. Bucketizing the actual values
allows us to extract the same information with a lower bit amount, which
would give the following mappings \{ 0=2043, 1=6833, 2=9700, 3=50900,
4=160000, 5=700000 \}. By using the bucket keys to encode the metric value
into the sort key, we still maintain the relative order between the actual
values whilst reducing the required amount of bits for encoding, in this
example within 3 bits. In case the bit amount is still too high for the
reserved bit range for a given metric, we aggregate the bucket values by
the standard deviation of the metric value range starting from the minimum
value. For example, the first bucket would cover values \(2043 + mult *
     \sigma\), where \(\sigma = 236988.01\). Having a lower \(mult=0.05\) value
allows us to capture \{ 0=\{ 2043,6833,9700 \}, 1=\{ 50900 \}, 2=\{ 160000 \}, 3=\{
700000 \} \} within 2 bits. As the entropy of the value range increases,
\(mult\) can be adjusted higher until the required bit amount covers the
whole metric value range.
\textbf{Sort Key encoding} We encode the metrics for each condition into a single
unsigned 32-bit integer to build \emph{sort keys}, which can be comfortably
sorted using the parallel sort function introduced in Section
\ref{sec:fork_join_sort_merge_instance}. The first 9 bits are allocated for
the \emph{number of inter-fact connections} (assumed to be at max. 512 bucket
keys). The next 11 bits can be used to store the \emph{estimated island score}
(max 2048 bucket keys). Then, 2 bits can be used to store the condition's
rank, since the available condition ranks are fixed at the value set
\(\{0,1,2,3\}\). The last 10 bits can be used to store the \emph{minimum rank 1
cardinality cost} for that condition (max 1024 bucket keys). The order of
the sorted conditions using such an encoding scheme follows the priorities:
\emph{number of inter-fact connections}, \emph{estimated island score}, \emph{condition
rank}, \emph{min. rank 1 cardinality cost}. As each rule's conditions can be
evaluated independently of each other, the thread parallel version of the
\emph{sort key} construction algorithm is simply running it inside separate
worker threads (see Algorithm \ref{algorithm:parallel_rule_eval_algorithm}).
Vectorization opportunities for \emph{sort keys} are not discussed in this
paper.

\subsubsection{Join-Level Structures and Algorithms}
\label{sec:org7d1a075}
\label{sec:join_structure_algorithms}

Rules are a composition of conditions that require a series of \texttt{RNL}
lookups. The \emph{rank 1 index} helps with the initial lookup of stored facts,
but the subsequent pattern matching on logical variables that are equal
among conditions (inter-condition joins) require another layer of data
structures to perform efficiently. For example, when joining the conditions
c1 = (\texttt{Person} ?p livesIn ?c \texttt{string}) and c2 = (\texttt{City}
?c name ``Vienna'' \texttt{string}) it would conceptually result in a
materialized result set of facts that includes all facts from the fact type
\texttt{Person}, where the \texttt{<value>} component is equal to the
\texttt{<id>} component of all facts that has type \texttt{City} and its
\texttt{<value>} component \(=\) ``Vienna''. This join happens due to the
common logical variable ?c on the respective components of both conditions.

In terms of storing these join results we can pursue the following designs:
(1) row-oriented, (2) column-oriented, and (3) bit vectors. In this paper we
focus on the first two approaches. The \emph{row-oriented} result data structure
holds a two-dimensional array of join results, where the inner array
represents rows of join results. The column of that array represents the
fields on which to slot in the join results. For example in conditions c1 and
c2, the fields would represent the logical variables ?p and ?c. The
\emph{column-oriented} data structure swaps rows with columns. Each inner row
holds all values of a single column. For example, the first index to the
first array of the \emph{column-oriented} two-dimensional array would hold all
values matching the logical variable ?p. One major advantage of this scheme
is compression, as the tightly packed inner array holds values of the same
value type and allows for techniques such as run-length encoding (RLE) and
delta encoding to be applied, as well as make better use of the CPU caches.
One major challenge compared to the \emph{row-oriented} design is the sorting of
facts.

By default, pattern matching for joins happens with an \texttt{EQUAL} check
on logical variables, which is not always applicable. In order to support
arbitrary joins based on custom boolean binary functions, we can define
\emph{variable join tests}.

\begin{definition}{{\small \textbf{[Variable Join Test (\texttt{I})]}}} We
  define a Variable Join Test to be a triple
  \vspace{-0.45cm}

   $$\texttt{(<var1> <operator> <var2>)}$$

  where \texttt{<var1>} refers to a logical variable inside the first
  condition and \texttt{<var2>} refers to a logical variable inside the
  second condition that the join operation is applied on. The \texttt{<value
  type>} of both conditions have to be the same in case the variables refer
  to the \texttt{<value>} component, and are internal string-id handles to
  the string dictionary otherwise. \texttt{<operator>} is a boolean binary
  function that takes \texttt{<var1>} and \texttt{<var2>} as input. Variable
  join tests come into play when joins between two conditions are performed,
  in which case the \texttt{<operator>} function is called and if passed, a
  successful pattern match is registered whereupon the facts are added to the
  join result data structure.

  \textbf{Example}: Given two conditions with a variable join test c1 =
   (\texttt{AgeClass} ?ac minAge ?acMin \texttt{uint32}), c2 =
   (\texttt{Person} ?p age ?pAge \texttt{uint32} [(?page $>=$ ?acMin)]) the
   join of these two conditions results in successful joins for facts of type
   \texttt{Person} and \texttt{AgeClass} where the person's age exceeds or
   matches the corresponding minium age class attribute.

\label{def:variable_join_tests}
\end{definition}

\textbf{Join Algorithms}. A plethora of join algorithms have been researched and
discussed
\cite{efa74e5dfb852d2fbb9a48862a427843,b913d50abea5e224196416290c4b1e5e,baeb5a24d59d88ccd391abfdcf2ae162}.
Hiperfact focuses on the basic hash join and the parallel sort merge join.
The latter has been discussed in existing research
\cite{efa74e5dfb852d2fbb9a48862a427843}. We pick up the discussion in the
context of fact triple databases, especially due to the underlying \emph{fork
join} model's applicability to other problems in the engine. It is not only
suitable as a join algorithm, but we can create instances of the model for
sorting (Section \ref{sec:island_processing}), filtering (as a uniqueness
filter in Section \ref{sec:derivation_tree_recursive}), and aggregating of
matched facts. The model is malleable for both thread as well as data
parallelization implementations in all of the mentioned instances. Due to
space constraints, we will limit the discussion to the join and sorting
instances of the \emph{fork join} model.

\textbf{Fork-Join Model}. The \emph{fork-join} model \cite{7548985} is composed of a \emph{fork}
function that splits the initial data into blocks, operates on the blocks
after the split, and finally joins those blocks together with a \emph{join}
function. The idea is that both thread and data parallelism shall be
employed. The former can be applied due to the blocks being able to be
processed independently per thread. Data parallelism can be -- depending on
the \emph{fork-join} instance -- employed in both the \emph{fork} function, as well as
the \emph{join} function. We now discuss each different instances of the
\emph{fork-join} model.

\begin{enumerate}
\item \textbf{Instance: Parallel Sort Merge for Sorting}.
\label{sec:orgeeb28eb}
\label{sec:fork_join_sort_merge_instance}

For read queries the parallel sort algorithm becomes a critical component as
it can be applied in several situations: (1) when sorting conditions within
a rule, (2) when sorting islands using sort keys (see Section
\ref{sec:island_processing}), and (3) when sorting intermediate join results
as well as the final matching facts. Figure \ref{parallel_sort_merge_overview}
shows a \emph{fork-join} model instance for a parallel sort merge operation.

\begin{figure}[htbp]
\centering
\includegraphics[width=9cm]{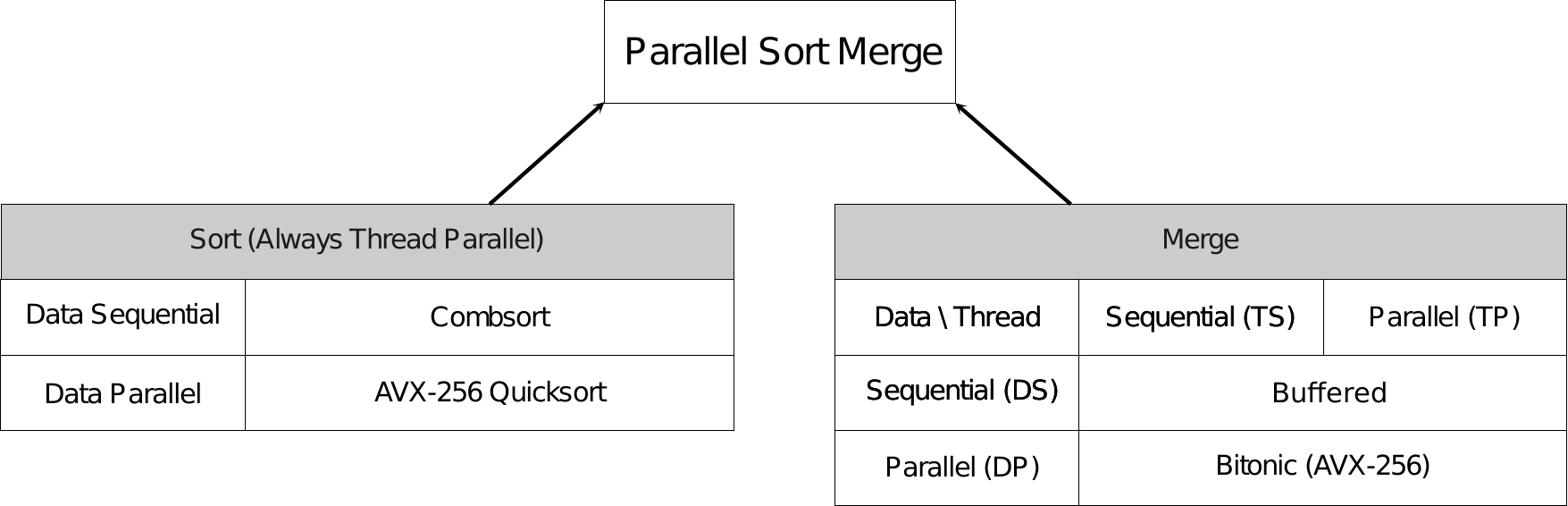}
\caption{\label{parallel_sort_merge_overview}Parallel Sort Merge Overview}
\end{figure}

\textbf{Fork Function}: For the implementation of the fork function for the sorting
problem, we have implemented a data parallel version based on AVX-256
Quicksort \cite{a7c876b59de9c3021c33a8b9c3350ff0} and a thread and data
sequential implementation of Combsort \cite{1c07684d12a395a56ede6eb1d63432a7}.
In this phase the algorithm focuses on partitioning the whole data set and
applies to those partitions the designated local sort algorithm. The whole
array is partitioned to be sorted according to a given block size. Jobs are
assigned to each partition for local sorting. To ensure spatial locality,
sub partitions are consecutively stored in tightly packed arrays. Each
thread owns the memory region allowing independent read and write access,
and thus ensuring concurrent sort of all blocks. A good first estimate for
the block size is the size of the L2 cache.

\textbf{Join Function}. After all partitions are sorted locally, each partition is
merged pairwise iteratively until no more partition pairs are found. During
the merging step, ensure that the sort order of the combined array is
maintained. To make this step parallel we merge two adjacent partitions per
thread. This individual merge thread can benefit from SIMD registers as
well. We take the AVX2-based merge operators (bitonic sort network) from
\cite{efa74e5dfb852d2fbb9a48862a427843} and adapt it for 32-bit integer values
as well. We also implement a sequential merge function for the basis of
calculating achieved speedup. To ensure no bottlenecks occur due to dynamic
memory management we pre-allocate the final sorted list twice for double
buffering. We alternate between those two buffers at each merge level such
that at no point do we require dynamic allocation for storing merge results.
Each thread has exclusive access to their memory ranges ensuring parallel
writes.

\item \textbf{Instances: Parallel Sort Merge Join, Unique Filter, Id+Object Sort}.
\label{sec:org34bd644}
\label{sec:join_algorithm_instance}

As mentioned before, the parallel sort merge join instance implements the
algorithm described in \cite{efa74e5dfb852d2fbb9a48862a427843}. It is
structured similar to the parallel sort instance, but since the input
consists of two arrays to be joined according to some join key, now each
thread holds a range partition, allowing local joins between the two locally
sorted fact arrays to be performed. Further instances include a parallel
sort unique filter, where the \emph{join} step performs a modified merge
operation filtering for unique values. This unique filter algorithm is used
primarily in \emph{derivation trees} (see Section \ref{sec:derivation_tree}) for
deduplication of inferred facts. Finally, a \emph{fork-join} instance for sorting
objects, where arbitrary objects are sorted according to their associated
integer \emph{id} by swapping the same index values in the corresponding arrays.
In this instance, the \emph{join} operation is identical to the parallel sort,
but the \emph{fork} operation includes the extra swap operations for the object
arrays. This \emph{fork-join} instance is heavily used for sort keys in Section
\ref{sec:sort_keys} and column-based join result structures, where the object
arrays correspond to the individual columns.
\end{enumerate}

\subsection{Derivation Trees}
\label{sec:orgf462eec}
\label{sec:derivation_tree}

\begin{figure}
\centering
\includegraphics[width=14cm]{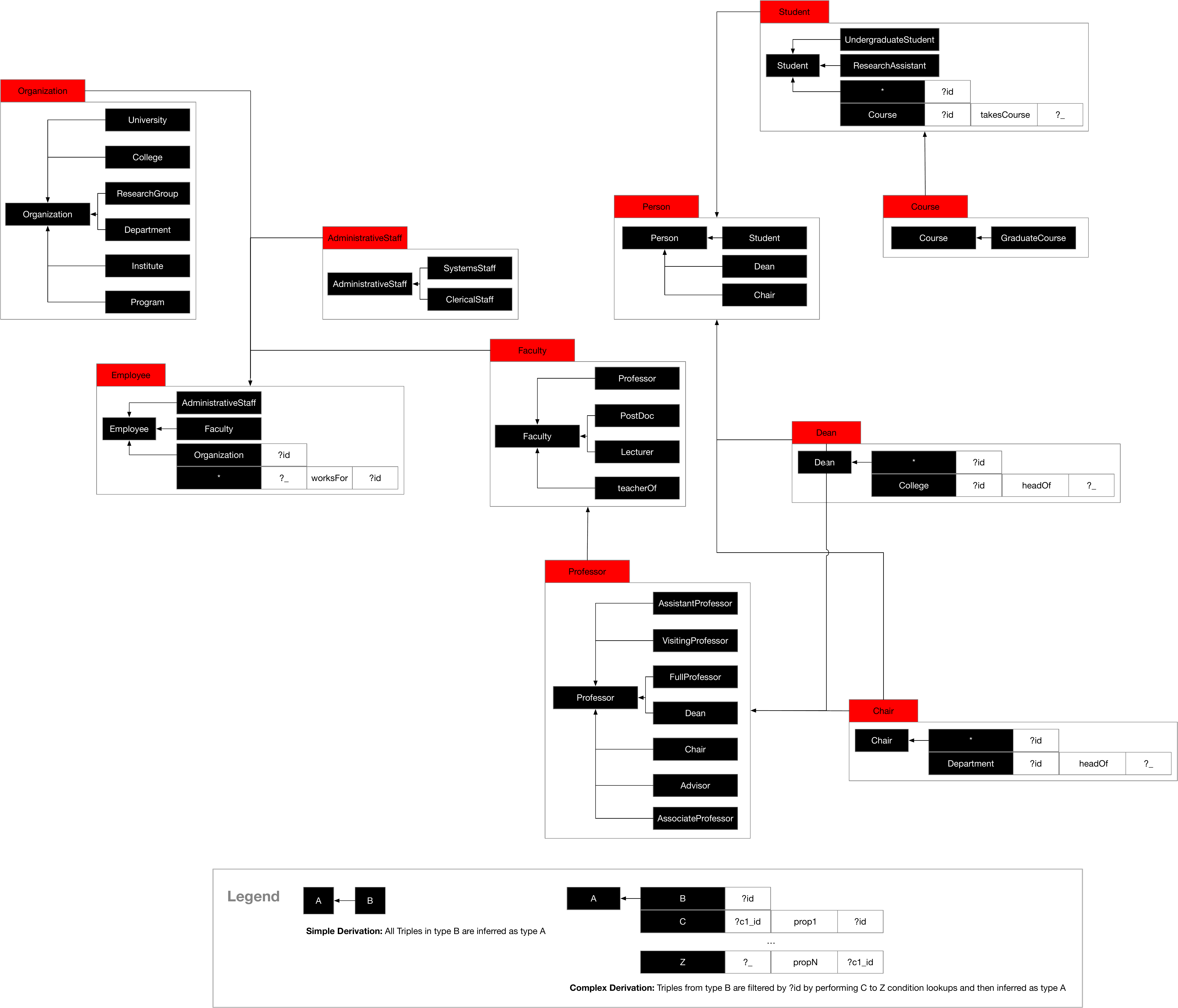}
\caption{Derivation Tree}
\label{fig:derivation_tree}
\end{figure}

The final component to discuss is the \emph{derivation tree} concept, dealing with
design chokepoints \emph{parallelism} and \emph{lazy evaluation} in the context of fact
inference. The motivation for this component is based on (\textbf{P2}), as having
the ability to filter out facts that are tied to \emph{inactive} rules is
beneficial for overall fact processing performance. In order to do so we need
to define the queries that are in effect: \emph{active} rules. We do this by
building a rule hierarchy that maps the fact type dependencies between the
rules. If the query does not necessitate certain types, then rules with those
types can be excluded from processing (passed to the \emph{island processing}
component). A derivation tree is based on the strong typing property of each
fact (see Definition \ref{def:fact}) in order to structure the flow of
inference (see Figure \ref{fig:derivation_tree}). Each node within this tree
maintains the resulting data type and maps to it all of the conditions that
map to that data type. The resulting data type is the \texttt{<fact type>} being
inserted in the action part of a rule.

\subsubsection{Parallel Execution using Derivation Trees}
\label{sec:orgb9d7b7b}
\label{sec:parallel_execution_derivation_tree}

One of the benefits the derivation tree can realize is the identification of
concurrent writing opportunities. Since the underlying data structure
responsible for storing facts are the rank 1 indices, combined with facts'
strongly typed property -- and thus being stored separately -- it becomes
possible to pipeline the index modification per \texttt{<fact type>}. The driving
factor is the \texttt{<fact type>}, which is a meta data associated to the rule
action. The top-down approach is used when processing derivation trees.
Starting from level 0 root nodes, representing the rules that have no
incoming edges (i.e. no rules are defined that result in derivation of this
fact type), rule processing is done by grouping them by their resulting fact
type, submitting each group to the job system for parallel execution. Each
group is then processed sequentially inside the assigned thread. By doing so
the rank 1 indices are modified consistently as each thread owns the memory
range associated to that fact type. The levels are followed and each rule
group is processed in the same fashion recursively until all levels are
traversed. Algorithm \ref{algorithm:parallel_rule_eval_algorithm} illustrates
this parallel execution method using derivation trees. Note that in Line 7
the sort keys are re-initialized after each derivation tree level, to take
into consideration any new cardinality updates due to any newly inferred
facts (or deletions). Thus after each update the join order is optimized,
given new data.

\begin{algorithm} [!h]
\begin{scriptsize}
\KwIn{}
\myinput{$deriv\_trees$ - \emph{derivation trees from all rules}}
\KwSty{Begin}\\

w $\leftarrow$ worker\_threads\_init(); \\
\ForEach{ $level$ in $deriv\_trees.levels()$ }
{
  \mc{ out\_group holds rules having write access to the same rank 1 index }
  \mc{ island construction (see Phase 1+2 from Algorithm \ref{algorithm:island_based_evaluation}) + associated sort\_key encoding }
  islands $\leftarrow$ build\_sort\_keys(level.rules()); \\
  sort(islands by sort\_keys grouped by rule); \\
  \ForEach{ $out\_group$ in $level.out\_groups()$ }
  {
    job $\leftarrow$ Job(); \\
    \ForEach{ $rule$ in $out\_group.rules()$ }
    {
      \mc{ island\_sort\_key\_eval is simliar to Algorithm \ref{algorithm:island_based_evaluation}, excluding Phase 1+2 and any sort logic on the islands array. islands[$rule$] holds the pre-sorted island array for that rule. }
      job.add(island\_sort\_key\_eval, islands[$rule$]);
    }
    w.add(job);
  }
  w.start();\\
  w.join();
}
w.delete();

\caption{Parallel Rule Evaluation (sort keys)}
\label{algorithm:parallel_rule_eval_algorithm}
\end{scriptsize}
\end{algorithm}
\vspace{-0.35cm}
\subsubsection{Lazy Rule Execution}
\label{sec:org5caee8e}
\label{sec:lazy_rule_eval_derivation_tree}

Drools \cite{06dde6ee870827728c0e31572574d9bb} has shown the underlying
problem of the forward chaining inference approach: all rules are followed
leading to fact derivations that are not always necessary, leading to time
wasted on unnecessary fact processing. Combined with Rete's memoizing
property, the problem is magnified due to caching unnecessary facts. To
tackle this issue, the derivation tree is used to identify rules that are
not required to be executed. In order to do so we distinguish rule nodes
between derivation rules and queries. The distinction lies in whether there
is an action associated to the rule that modifies facts. The existence of
such actions signals the rule to be a derivation rule, and a query
otherwise. Stated differently, leaf nodes in the derivation trees represent
queries, and nodes with children represent derivation rules. The idea is to
pre-define together with the flow of derivation rules the queries that are
intended to be executed. These usually represent standard queries that are
necessary for already known analysis tasks. These rules are then put
together to create the derivation trees. When new queries are added while
the Hiperfact system is running (e.g. ad-hoc queries), inactive derivation
rules may become active to cover the new query. This process happens
naturally as the derivation trees are rebuilt any time rules are modified.
Following the derivation trees in a bottom-up fashion starting from leaf
nodes (queries) allows us to identify inactive rules (see Definition
\ref{def:active_rule}). Derivation Rules where \(RT(r) =
    \texttt{DERIVATION\_RULE} \wedge AR(r) = \texttt{false}\) can be skipped for
processing as they do not depend on any queries.
\begin{definition}{{\small \textbf{[Rule Type]}}} We define the rule
  type function \texttt{RT} accepting a rule as parameter (see Definition
  \ref{def:rule}) as:

 \[ \texttt{RT}(r) = \begin{cases} \texttt{DERIVATION\_RULE}, & \text{if } \texttt{children}(r) > 0\\ \texttt{QUERY}, & otherwise \end{cases} \]

 \texttt{children} is a function returning the number of rule nodes that
 will be executed next according to the derivation tree. A rule node $c$ is
 a child of another rule $p$ when $c$ holds a \\ \texttt{<fact type>} in any
 of its conditions that is marked being modified in $p.action$.

\label{def:rule_type}
\end{definition}
\begin{definition}{{\small \textbf{[Active Rule]}}} To determine
  whether a given Rule r is active, and thus needs to be evaluated, we
  define \texttt{AR} accepting, as parameter, a rule as follows:
  \vspace{-0.35cm}

 \[ \texttt{AR}(r) = \begin{cases} \texttt{true}, & \exists x \in children(r), \texttt{AR}(x) \vee \texttt{RT}(x) = \texttt{QUERY} \\ \texttt{false}, & otherwise \end{cases} \]

 A rule is marked as active when a rule of type QUERY is on its path (i.e.
 from the leaf to the root node).

\label{def:active_rule}
\end{definition}

\subsubsection{Recursive Execution of Derivation Trees}
\label{sec:org731bd24}
\label{sec:derivation_tree_recursive}

As \emph{derivation trees} could by cyclic, where root nodes link to children
nodes, the actual execution outlined in Section
\ref{sec:parallel_execution_derivation_tree} needs to be repeated until a
fixpoint is reached, i.e., where no new facts are inferred. Doing so might
lead to facts that have been previously inferred. As such, an efficient
deduplication operation of facts needs to be employed to be viable. We use
the parallel unique filter discussed in Section
\ref{sec:join_algorithm_instance}.

\section{Experimental Evaluation}
\label{sec:org759b2e8}
\label{sec:evaluation}

In this section we discuss the experimental evaluation to observe the
effectiveness of the introduced components and thus also the Hiperfact
architecture. The primary measurements are the run-time performance metrics:
data set loading time, fact inference time, and query time. Secondary measures
are tracked as well for observing the last level (LL) and L1 CPU cache miss
rate and the instructions per cycle. A high number of instructions executed
per cycle, combined with a lower cache miss rate is indicative of a
well-performing implementation, but these secondary measurements are tracked
over the whole implementation and cannot pinpoint the individual components'
performance in that regard. For the primary metrics, we can show the complete
breakdown of the total run-time.

\begin{figure}
\centering
\includegraphics[width=8cm]{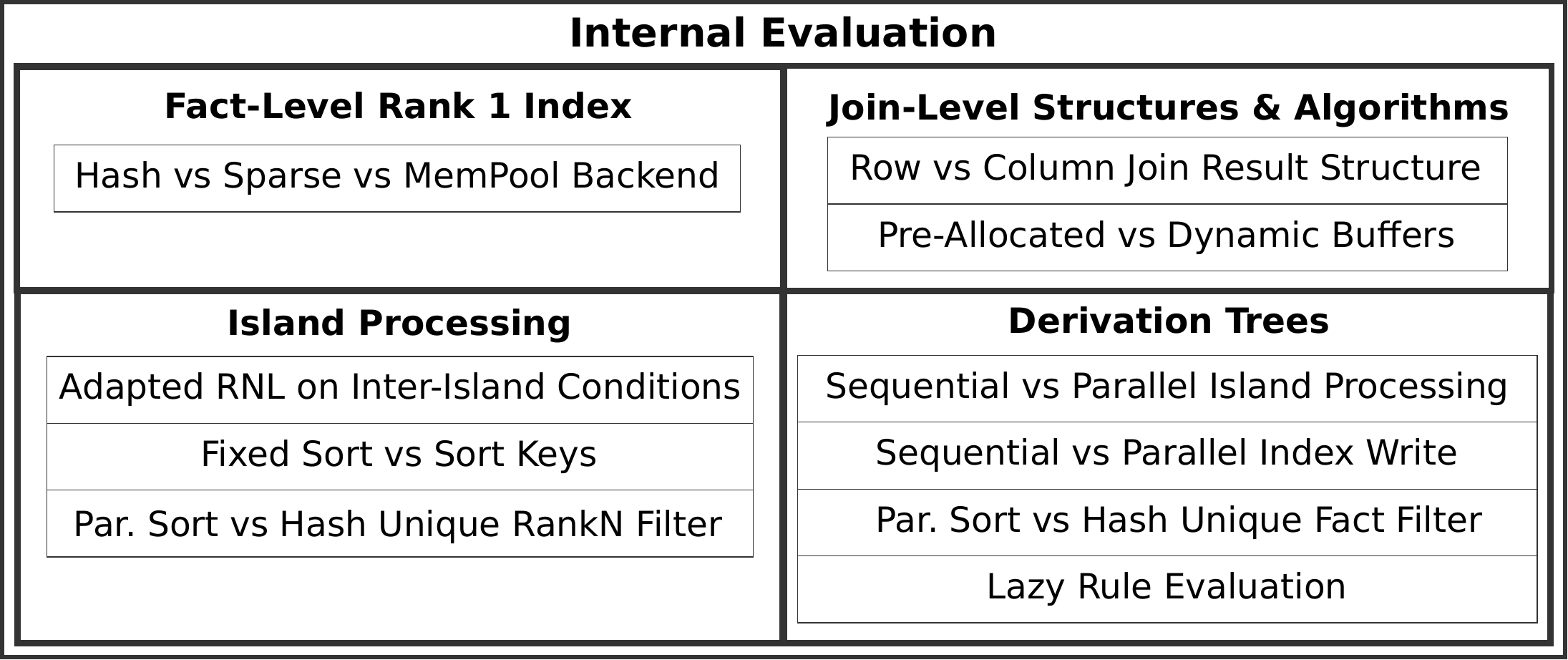}
\caption{Component Evaluation Overview}
\label{evaluation_components}
\vspace{-0.35cm}
\end{figure}

Figure \ref{evaluation_components} shows what aspects within the individual
Hiperfact engine components are configurable and thus affect the final
performance. Table \ref{tab:benchmark_acronym_table} highlights the acronyms
each of those configurations are mapped to, which will be referred to from
this point on. For the \emph{rank 1 index}, notable configurations are the
different backend implementations (AI,HI,LPIM,LPID) responsible for fact
storage and retrieval, which have been touched on in Section
\ref{sec:rank1_index}.

For the query-focused \emph{island processing} we have different join algorithms:
hash join (HJ) and parallel sort merge join (MJ), briefly mentioned in Section
\ref{sec:join_structure_algorithms}. For performing the inter fact join \emph{rank 1
index} lookup (RNL) there are two: (AR) performing a rank \(<\) 3 lookup for
each inter-island condition with variables mapped from the now known \emph{join
results} table before joining with the target island and (DR) which is the
default RNL operation without consideration of the existing \emph{join results} to
join with the next island. Furthermore, for the \emph{island processing} component,
we distinguish between a columnar (CR) and row-based (RR) data structure for
storing intermediate join results. For the inference-focused \emph{derivation tree}
component, we have configurations for parallel (PR) vs sequential (SF)
\emph{derivation tree node} processing, and the related parallel (PW) vs sequential
(SW) index write approaches. These have been discussed in Section
\ref{sec:parallel_execution_derivation_tree}. Finally, configurations exist for
applying a unique filter on newly inferred facts, which might be already
inferred at a previous inference loop. To discover such occurrences we employ
a parallel sort merge based (SU) vs a hash table based (HU) unique filter.
\begin{table}[h]
\begin{scriptsize}
\centering
\ra{1.2}
\begin{tabular}{|c|c|}\toprule
Acronym & Description \\ \midrule
\textbf{Rank 1 Index} & \\
\,\,\,\, AI   & 3-level Sparse Array Index \\
\,\,\,\, HI   & Hashtable Index \\
\,\,\,\, LPIM & Array Linked Pages Index + Memory Pool \\
\,\,\,\, LPID & Array Linked Pages Index + Dynamic Memory \\
\textbf{Island Processing} & \\
\,\,\,\, HJ & Inter/Intra-Fact Hash Join \\
\,\,\,\, MJ & Inter/Intra-Fact Parallel Sort Merge Join \\
\,\,\,\, AR & Adapted RNL \\
\,\,\,\, DR & Default RNL \\
\,\,\,\, CR & Columnar Intermediate Join Results \\
\,\,\,\, RR & Row-based Intermediate Join Results \\
\textbf{Derivation Trees} & \\
\,\,\,\, PF & Parallel Node Level Queries \\
\,\,\,\, SF & Sequential Node Level Queries \\
\,\,\,\, PW & Parallel Index Write \\
\,\,\,\, SW & Sequential Index Write \\
\,\,\,\, SU & Parallel Sort Merge Unique Filter \\
\,\,\,\, HU & Hashtable Incremental Unique Filter \\ \midrule
\textbf{Configurations} & \\
\,\,\,\, $\text{Hiperfact}_{\texttt{infer1}}$  & LPIM+HJ/AR/CR+PF/PW/SU \\
\,\,\,\, $\text{Hiperfact}_{\texttt{query1}}$  & AI+MJ/AR/CR+PF/PW/SU \\
\bottomrule
\end{tabular}

\caption{Hiperfact configurations}

\label{tab:benchmark_acronym_table}
\vspace{-0.7cm}
\end{scriptsize}
\end{table}

From several runs we have two pre-configured instances of Hiperfact: \emph{infer1}
and \emph{query1}. The former focuses on inference workloads and uses a memory
pooled array linked pages (LPIM) \emph{rank 1 index}, hash joins (HJ) for the join
operation using the adapted RNL (AR) approach with columnar join result
storage (CR) for \emph{island processing}, and parallel \emph{derivation tree} node
processing for read (PF), parallel index writing (PW) together with a parallel
sort merge unique filter (SU) for uniqueness checks. The \emph{query1} preset
differs in that the \emph{rank 1 index} backend employs the three-level sparse
array index (AI instead of LPIM).
\textbf{Experimental Setup}. For the experiments we used different machine setups.
Due to space reasons, we will only include the inference and query
benchmarks, for which we used a physical system running an AMD Ryzen 9 2950X
@3.5GHz (16 Cores) CPU with 64GB of RAM having L1 and L2 cache sizes of 512KB
and 8MB respectively.

\subsection{Benchmarks}
\label{sec:org0449c1f}

With the configurations defined, we run both inference focused benchmarks
based on LUBM \cite{lubm} and WordNet \cite{wordnet}, as well as query focused
benchmarks as specified in OpenRuleBench \cite{openrulebench}. We also run
existing engines to compare the run-time performance. Each engine+data set
pair is executed 4 times, where the primary and secondary metrics are
tracked. The secondary \emph{perf} metrics: instructions per cycle, LL cache miss
rate, and L1 Data cache miss rate are tracked using the Linux \emph{perf} tool. We
verify that each engine's run-time performance is not negatively affected due
to monitoring under \emph{perf} by an additional run of that engine+data set pair
outside of \emph{perf}. In the case where a benchmark script exists (Inferray), we
explicitly use the command that script calls and execute it under \emph{perf} to
track the metrics. The first run is not tracked, allowing the caches to be
primed, and the subsequent 3 runs are then tracked and the average metrics
are reported in tables \ref{tab:benchmark_inference_times},
\ref{tab:benchmark_inference_perf_metrics} for inference and tables
\ref{tab:benchmark_query_times}, \ref{tab:benchmark_query_perf_metrics} for query
focused benchmarks respectively.
\begin{table*}[h]
\begin{scriptsize}
\centering \ra{1.2}
\begin{tabular}{@{}lcccccccc@{}}\toprule
& & Time Elapsed (in sec) & \\
& Load & Inference & Query \\
\midrule
\textbf{LUBM1}\\
\,\,\,\, RDFox               & 0.101 & 0.038 & \bu{0.002} \\
\,\,\,\, Inferray            & 0.42       & 0.168      &  N/A       \\
\,\,\,\, $\text{Hiperfact}_{\texttt{infer1}}$ & 0.544      & 0.0398     & 0.0021     \\
\,\,\,\, $\text{Hiperfact}_{\texttt{query1}}$ & \bu{0.0542} & \bu{0.0295} & \bu{0.0015} \\
\midrule
\textbf{LUBM50}\\
\,\,\,\, RDFox               & 5.453    & \bu{0.646} &	0.152      \\
\,\,\,\, Inferray            & 22.959   & 1.632      &	N/A        \\
\,\,\,\, $\text{Hiperfact}_{\texttt{infer1}}$ & \bu{3.9} & 0.765      & 0.104 \\
\,\,\,\, $\text{Hiperfact}_{\texttt{query1}}$ & 3.485 & 0.829 & \bu{0.082} \\

\midrule
\textbf{LUBM100}\\
\,\,\,\, RDFox               & 9.210       & \bu{1.291} & 0.314	    \\
\,\,\,\, Inferray            & 40.645	    & 3.150      & N/A	      \\
\,\,\,\, $\text{Hiperfact}_{\texttt{infer1}}$ & 7.342	      & 1.448	     & 0.200    	\\
\,\,\,\, $\text{Hiperfact}_{\texttt{query1}}$   & \bu{6.839}	& 1.573	     & \bu{0.149}	\\
\,\,\,\, $\text{Hiperfact}_{\texttt{infer1}}$+HU   & 7.412	& 3.363	     & 0.203	\\
\,\,\,\, Hiperfact (HI+HJ/DR/RR+SF/SW/HU)  & 7.778	& 13.491 & 0.228	\\
\midrule
\textbf{WordNet}\\
\,\,\,\, RDFox               & \bu{1.093}       & 0.2378 & N/A	    \\
\,\,\,\, Inferray            & 4.126	    & 0.475      & N/A	      \\
\,\,\,\, $\text{Hiperfact}_{\texttt{infer1}}$ & 3.75	      & \bu{0.1318}	     & N/A    	\\
\bottomrule \\
\end{tabular}

\caption{LUBM Inference+Query Time Elapsed (Average Values of 3 Runs)}

\label{tab:benchmark_inference_times}
\vspace{-0.5cm}
\end{scriptsize}
\end{table*}

\begin{table*}[h]
\begin{scriptsize}
\centering \ra{1.2}
\begin{tabular}{@{}lcccccccc@{}}\toprule
& & \texttt{perf} metrics & & \\
& Instr./cycle & LL Cache Miss(\%) & L1D Cache Miss(\%) \\
\midrule
\textbf{LUBM1}\\
\,\,\,\, RDFox               & \bu{1.823} & 12.39\%      & \bu{0.81\%}    \\
\,\,\,\, Inferray            & 1.20  & 20.14\%      & 5.17\%   \\
\,\,\,\, $\text{Hiperfact}_{\texttt{infer1}}$ & 1.213 & 10.96\% & 6.46\% \\
\,\,\,\, $\text{Hiperfact}_{\texttt{query1}}$ & 1.62 & \bu{8.295\%} & 5.42\% \\
\midrule
\textbf{LUBM50}\\
\,\,\,\, RDFox               & 1.543      & 12.31\%      & 3.49\% \\
\,\,\,\, Inferray            & 1.626      &	13.26\%	     & \bu{2.10\%} \\
\,\,\,\, $\text{Hiperfact}_{\texttt{infer1}}$ & \bu{2.013}	& \bu{10.52\%} & 3.31\% \\
\,\,\,\, $\text{Hiperfact}_{\texttt{query1}}$ & 1.61 & 12.904\% & 3.15\% \\

\midrule
\textbf{LUBM100}\\
\,\,\,\, RDFox               & 1.353      &	13.38\%	    &	3.77\% \\
\,\,\,\, Inferray            & 1.506      &	13.50\%	    & \bu{2.11\%} \\
\,\,\,\, $\text{Hiperfact}_{\texttt{infer1}}$ & 1.796      & \bu{8.58\%}  & 2.92\% \\
\,\,\,\, $\text{Hiperfact}_{\texttt{query1}}$   & \bu{1.892} & 11.05\%      & 2.88\% \\
\,\,\,\, $\text{Hiperfact}_{\texttt{infer1}}$+HU   & 1.74 & 18.412\%      & 2.91\% \\
\,\,\,\, Hiperfact (HI+HJ/DR/RR+SF/SW/HU)  & 2.68 & 4.083\%      & 20.98\% \\
\midrule
\textbf{WordNet}\\
\,\,\,\, RDFox               & N/A      &	N/A    &	N/A \\
\,\,\,\, Inferray            & \bu{1.46}      &	16.81\%	    & \bu{2.88\%} \\
\,\,\,\, $\text{Hiperfact}_{\texttt{infer1}}$ & 1.15      & \bu{16.44\%}  & 5.17\% \\
\bottomrule \\
\end{tabular}

\caption{LUBM Inference+Query \texttt{perf} metrics (Average Values of 3 Runs)}

\label{tab:benchmark_inference_perf_metrics}
\vspace{-0.5cm}
\end{scriptsize}
\end{table*}

\textbf{Inference Benchmarks (cf. Table \ref{tab:benchmark_inference_times} and Table
\ref{tab:benchmark_inference_perf_metrics})} The LUBM benchmarks are scaled
from 100k (LUBM1), 10M (LUBM50) to 17M (LUBM100) triples for the initial
data set. WordNet has a data size of nearly 2M triples. Using the RDFS-Plus
rule set, new facts are inferred before the systems are ready to be queried
using the query template provided by the LUBM benchmark. We compare our
engine against two inference engines: Inferray
\cite{8f2df16f0f940d37bd677bd56e5ae115} and RDFox \cite{rdfox}. The former is
specialized for handling RDFS-Plus rulesets and is thus hard coded into the
implementation, the latter allows the definition of the ruleset using a
datalog-like language as input. As RDFS-Plus can be seen as meta rules that
initializes the concrete rules from the given fact types, we have
implemented those instantiated rules directly as rules for the Hiperfact
engine. We argue that this is an acceptable approach allowing us to proceed
with the evaluation without having a concrete language parser and rule
generator in place. The focus of this evaluation is how fast these rules are
executed for the purpose of fact inference. We plan to add a Datalog parser
at a later point.

As for the overall result, Hiperfact performs well against the compared
systems. Regarding the time for loading triples, the sparse array index (AI)
consistently performs better than the array linked page indices (LPIM, LPID)
and scales better compared to RDFox and Inferray. The load performance
anomaly in the WordNet data set can be explained due to the reverse
declaration of types. Whereas the data type is usually declared first for a
given object, in WordNet, this is done in the end, thus requiring a
buffering process to hold the incomplete fact while the type is yet to be
declared. The extra memory allocation required to do so is reflected in the
lower loading time for this specific data set. One approach to improve this
behaviour is to introduce a memory pool of pre-allocated triples to hold
those temporary triples. In regards to inference performance, RDFox's
\cite{rdfox} concurrent core data structure becomes evident, as the increase
in number of facts to be inferred widens the gap in performance in their
favor compared to Inferray and Hiperfact. Hiperfact's current \emph{rank 1
indices} do not have a concurrent core implementation which would allow full
write access to the index from multiple worker threads. Despite that, we
were still able to exploit several parallelization points implemented via PF
and PW. In PF we used the \emph{derivation tree} to identify which nodes are
independent of each other, allowing parallel querying of rule conditions
leading to positive rule activations. In PW we then group these facts by
their data type to be processed concurrently, as the indices allow
concurrent write access per data type. This is reflected in the good scaling
behaviour, similar to RDFox's but slightly worse, using the \texttt{infer1}
configuration tuned for inference.

For Inferray, we were not able to reproduce their reported
\cite{8f2df16f0f940d37bd677bd56e5ae115} benchmark behaviour, using the
benchmark supplied by the
authors\footnote{\url{https://github.com/telecom-se/USE-RB/tree/77b8544a48e53181619e3f2a26b1db84ba5b2b6d}}\textsuperscript{,}\,\footnote{\url{https://github.com/telecom-se/ReasonersBenchmarked/tree/b033321c9aded71f06a2bffb8fa31d88550f2a3a}}. The reason seems to
be the inaccurate reporting of the benchmark numbers when running the
Inferray engine using their benchmark framework, compared with the numbers
extracted from the log that is generated when the Inferray engine is
executed. By default, 5 iterations are executed and the average of those 5
runs is claimed to be reported. By investigating the logs, we came to the
conclusion that the best of the 5 runs are reported, which is always the
last run. Furthermore, as the iteration number increases, the loop time
improves by a consistent 10\% to 38\% depending on the data set. And since
only the last iteration counts, the best time is being logged for the
benchmark, which is misleading as in order to achieve that number, the
effort of the previous 4 iterations needs to be included as well. Due to the
consistent improvement in loop time after each iteration, even an average of
those numbers leads to an inaccurate measure of total effort. Since no other
engine under evaluation behaves this way (improved time after subsequent
iteration), combined with the fact that we need to run all engines under
\emph{perf} to track secondary measures, and that Inferray itself does not show
this incremental improvement behaviour when run with iteration number of 1
but 5 times, we do exactly that: run the benchmark program with an iteration
number of 1 and follow the tracking methodology outlined in the beginning of
this section using the reported inference times from the log Inferray
generates.
\begin{table*}[h]
\begin{scriptsize}
\centering
\ra{1.2}
\begin{tabular}{@{}lccccccc@{}}\toprule
& Time Elapsed (in sec) &\\
& Load & Query & \\
\midrule
\textbf{Mondial}\\
\,\,\,\, RDFox                     & 0.133	     & 2.704         \\
\,\,\,\, Hiperfact (LPIM+HJ/AR/CR) & 0.155      & 0.000418      \\
\,\,\,\, Hiperfact (LPIM+HJ/DR/CR) & 0.154      & 0.00159       \\
\,\,\,\, Hiperfact (LPIM+HJ/AR/RR) & 0.156      & 0.000575      \\
\,\,\,\, Hiperfact (LPIM+MJ/AR/CR) & 0.154      & \bu{0.000401} \\
\,\,\,\, Hiperfact (LPID+HJ/AR/CR) & \bu{0.125} & 0.000482      \\
\,\,\,\, Hiperfact (AI+HJ/AR/CR)   & \bu{0.126} & \bu{0.000395} \\
\,\,\,\, Hiperfact (AI+MJ/AR/CR)   & \bu{0.127} & \bu{0.000354} \\
\,\,\,\, Hiperfact (AI+HJ/AR/RR)   & \bu{0.123} & 0.000569      \\
\,\,\,\, Hiperfact (AI+HJ/DR/CR)   & 0.130	     & 0.000900      \\
\midrule
\textbf{DBLP}\\
\,\,\,\, RDFox                     & \bu{0.820} & 0.033	     \\
\,\,\,\, Drools                    & 1.887      & 1.116       \\
\,\,\,\, XSB                       & 40.149	   & 1.126	     \\

\,\,\,\, Hiperfact (LPIM+HJ/AR/CR) & 1.458			 & 0.0018	           \\
\,\,\,\, Hiperfact (LPIM+HJ/DR/CR) & 1.435			 & 0.0231	           \\
\,\,\,\, Hiperfact (LPIM+HJ/AR/RR) & 1.522			 & 0.0027			 \\
\,\,\,\, Hiperfact (LPIM+MJ/AR/CR) & 1.505			 & 0.0016			       \\
\,\,\,\, Hiperfact (LPID+HJ/AR/CR) & 1.453			 & 0.0024			       \\
\,\,\,\, Hiperfact (AI+HJ/AR/CR)   & 1.403			 & 0.0016			       \\
\,\,\,\, Hiperfact (AI+MJ/AR/CR)   & 1.506			 & \bu{0.0013} \\
\,\,\,\, Hiperfact (AI+HJ/AR/RR)   & 1.476			 & 0.0024			 \\
\,\,\,\, Hiperfact (AI+HJ/DR/CR)   & 1.425			 & 0.0093			 \\
\,\,\,\, Hiperfact (AI+MJ/DR/CR)   & 1.505		   & 0.0241			 \\

\bottomrule \\
\end{tabular}

\caption{OpenRuleBench Query Time Elapsed Results (Average Values of 3 Runs)}

\label{tab:benchmark_query_times}
\vspace{-0.5cm}
\end{scriptsize}
\end{table*}

\begin{table*}[h]
\begin{scriptsize}
\centering
\ra{1.2}
\begin{tabular}{@{}lccccccc@{}}\toprule
& & \texttt{perf} metrics & \\
& Instr./cycle & LL Cache Miss(\%) & L1D. Cache Miss(\%) \\
\midrule
\textbf{Mondial}\\
\,\,\,\, RDFox                     & \bu{1.633} & 20.48\%  & 8.62\%       \\
\,\,\,\, Hiperfact (LPIM+HJ/AR/CR) & 1.353 & \bu{10.62\%}  & \bu{1.42\%}  \\
\,\,\,\, Hiperfact (LPIM+HJ/DR/CR) & 1.373 & \bu{9.70\%}	 & \bu{1.45\%} \\
\,\,\,\, Hiperfact (LPIM+HJ/AR/RR) & 1.386 & \bu{10.24\%}  & \bu{1.57\%}  \\
\,\,\,\, Hiperfact (LPIM+MJ/AR/CR) & 1.433 & \bu{10.28\%}  & \bu{1.52\%}  \\
\,\,\,\, Hiperfact (LPID+HJ/AR/CR) & 1.273 & \bu{10.96\%}  & \bu{1.49\%}  \\
\,\,\,\, Hiperfact (AI+HJ/AR/CR)   & 0.753 & 26.80\% & 1.65\% \\
\,\,\,\, Hiperfact (AI+MJ/AR/CR)   & 0.793 & 27.20\% & 1.65\% \\
\,\,\,\, Hiperfact (AI+HJ/AR/RR)   & 0.863 & 19.87\% & 1.64\% \\
\,\,\,\, Hiperfact (AI+HJ/DR/CR)   & 0.683 & 34.62\% & 1.70\% \\
\midrule
\textbf{DBLP}\\
\,\,\,\, RDFox                     & 1.033 & 22.53\% & \bu{2.31\%} \\
\,\,\,\, Drools                    & 1.086 & 22.98\% & 4.88\% \\
\,\,\,\, XSB                       & \bu{1.48}	 & \bu{3.98\%}	 & 3.86\% \\

\,\,\,\, Hiperfact (LPIM+HJ/AR/CR) & 1.463 & 21.26\%	     & 3.29\% \\
\,\,\,\, Hiperfact (LPIM+HJ/DR/CR) & 1.41	 & 21.19\%			 & 3.31\% \\
\,\,\,\, Hiperfact (LPIM+HJ/AR/RR) & \bu{1.48}	 & 21.18\%			 & 3.06\% \\
\,\,\,\, Hiperfact (LPIM+MJ/AR/CR) & 1.47	 & 20.63\%			 & 3.00\% \\
\,\,\,\, Hiperfact (LPID+HJ/AR/CR) & 1.32	 & 21.87\%			 & 3.42\% \\
\,\,\,\, Hiperfact (AI+HJ/AR/CR)   & 1.23	 & 26.89\%			 & 3.86\% \\
\,\,\,\, Hiperfact (AI+MJ/AR/CR)   & 1.24  & 26.40\%			 & 3.70\% \\
\,\,\,\, Hiperfact (AI+HJ/AR/RR)   & 1.23	 & 27.34\%			 & 3.61\% \\
\,\,\,\, Hiperfact (AI+HJ/DR/CR)   & 1.25	 & 27.12\%			 & 3.69\% \\
\,\,\,\, Hiperfact (AI+MJ/DR/CR)   & 1.38	 & 22.83\%			 & 3.97\%	\\

\bottomrule \\
\end{tabular}

\caption{OpenRuleBench Query \texttt{perf} metrics (Average Values of 3 Runs)}

\label{tab:benchmark_query_perf_metrics}
\vspace{-0.5cm}
\end{scriptsize}
\end{table*}

\textbf{Query Benchmarks (cf. Table \ref{tab:benchmark_query_times} and Table
\ref{tab:benchmark_query_perf_metrics})} In terms of loading time, dynamic
memory allocation configurations (AI,LPID) win due to the overhead of
pre-allocating memory pools before they are required. Default RNL (DR)
Lookup in all configurations are not interesting in terms of island
processing as unnecessary facts are fetched that are joined during intra-
as well as inter-fact joins. The adapted RNL fetches individual facts to
be joined based on the actual values already in the join result buffer.
The sparse array index is preferred for querying as Rank 1 Lookups (R1L)
can simply take the pointer to the index holding required facts without
having to copy elements around. RDFox seems to stall the least and
exhibits the highest instructions per cycle. In this regard the Linked
pages Index (LPIM/LPID) fare better compared to the sparse array index
(AI) both in terms of less stalled instructions and better L2/L1 cache
utilization. Having pre-allocated pages ready to be claimed by the index
is beneficial for island processing. Another point is that less abrupt
calls to dynamic memory allocation are needed leading to less copying of
facts to accomodate the bigger buffer for holding intermediate results.
The degraded query performance for RDFox in the \emph{mondial} data set seems
to be due to a cross product between the two classes to be joined. The
used academic license version of RDFox might not have a full SPARQL query
optimizer to perform efficient joins. XSB uses the L2 cache the most
efficient. Regarding the difference in RNL lookup variations (AR vs DR),
it is clear that AR is advantageous when the join result leads to a
smaller number of lookups, which is the case for the \emph{mondial} and \emph{dblp}
data sets. In the case where AR is not selective, then the individual
rank \(<\) 3 lookups will perform worse than a singular RNL lookup before
the actual join, due to non-efficient usage of the CPU caches as multiple
random memory locations are visited (in the AR case) compared to one
consecutive array to be filtered (in the DR case). In conclusion, query
execution performance is best achieved using a three-level sparse array
index (AI) for \emph{rank 1 indices}, combined with a parallel sort merge join
(MJ), adapted RNL (AR) and column-oriented join result (CR) data
structure in the \emph{island processing} component.

\section{Related Work}
\label{sec:org4d8feb4}

Inferray \cite{8f2df16f0f940d37bd677bd56e5ae115} follows the general forward
chaining inference approach from Rete and addresses problem (4) through the
use of tightly packed arrays and problem (1) by never storing intermediate
results. Since the focus of Inferray is to perform inference in the Resource
Description Framework (RDF) context, it implements specialized rules to solely
cover that domain. In most cases, these rules handle equivalence between fact
types, neither requiring complex pattern matching logic involving any of
LIMIT, ORDER BY, JOIN nor aggregation functions. Another aspect to consider is
the intended workflow when employing Inferray. The design is to perform the
inference with the complete fact set preloaded and after materializing all
inferred facts subsequently executing queries on the final data set. Any newly
added facts after materialization will trigger a complete inference process of
the whole data set from scratch. This limits interactive exploration of the
data set.

Stylus \cite{91fd3608b3799a815471ec5708af5f6f} focuses on improved pattern
matching performance (of SPARQL queries) by introducing strong typing
capabilities during the query optimization process. It works independently of
any inference approach and solely focuses on query execution. The assumption
here is that the inferred facts are all pre-materialized on which the Stylus
algorithm then runs the queries. It claims to use the Inferray approach for
that purpose.

\cite{91fd3608b3799a815471ec5708af5f6f} divides facts by their entity type,
avoiding unnecessary joins. In our work, we extend the triple structure by
additional dimensions, namely to namespace facts via strong typing (see
Definition \ref{def:fact}). Only facts within the explicitly queried namespace
are candidates for joining. Additionally, their approach converts all triple
components independent of their type to an internal id for processing. In
contrast, in our approach we further distinguish each data type and their
relevant natural compression technique.

\cite{b260dee9ff2be53e2e7d030c020d45f9} analyses SPARQL queries and performs
join re-ordering by analysing characteristic pairs to identify foreign
relationships (one-to-many, many-to-many). This information is used to gain
better join result estimation and order the stars comprising the query. We do
not use characteristic pairs to identify foreign relationships, but use the
underlying index information (\emph{rank-1 indices}), and calculate the aggregated
sums for conditions and finally islands to get good lookup and join estimates.
The islands are processed in that order to keep the intermediate results as
small as possible. Furthermore we use thread as well as data parallelism to
further increase the performance of joins. The concept of stars in this work
corresponds to the islands in our approach.

\cite{20201118205457} focuses on the querying aspect and performs a query
optimization step for structuring fact processing in an optimized manner.
Similar to our island processing approach, it identifies islands that can be
processed independently and processes these islands in parallel using a mixed
CPU-GPU approach with the cost of maintaining all island results in memory
before performing a final join. Following the \ref{tpc_island_based_inference},
the islands are identified as: \texttt{S1 = (((STORE\_SALES JOIN CUSTOMER) JOIN DATE)
  JOIN INVENTORY)}, \texttt{S2 = (STORE\_RETURNS JOIN CUSTOMER)}, \texttt{S3 = (CATALOG\_SALES
  JOIN CUSTOMER)}.

\section{Conclusion and Future Work}
\label{sec:org6d8160e}

In this paper, we introduced the Hiperfact architecture, its components and
algorithms that deal with both inference and querying of triple structured
facts, that is common in RDF. The \emph{derivation tree} component tackles the
challenges evident in inference tasks: deduplication of inferred facts that
already exist through the use of an optimized data and thread parallel sort
and merge unique filter, lazy evaluation of inference rules that are only
processed when associated queries exist. The \emph{island processing} component,
combined with the \emph{rank 1 index} component and \emph{join result} data structures
and algorithms tackle the challenge of query optimization for fast retrieval
of facts. This is achieved by implementing different \emph{rank 1 index} backends
following a common API to observe the behaviour of different array-based fact
indices as well as implementing row-based and column-based intermediate join
structures. The main challenge tackled in the \emph{island processing} component is
finding an efficient join order through the use of cost estimation derived
from cardinalities in the \emph{rank 1 index}. Furthermore the strong typing of
facts, allows for natural parallelization for write and read access in the
\emph{derivation tree} component, as well as in the \emph{island processing} component.
All components strive to use the CPU caches (L1/L2) as efficiently as possible
by proper aligning of fact data in the \emph{rank 1 index} component and the \emph{join
result} data structures.

Hiperfact is slightly less performant regarding inference compared to RDFox,
but comfortably beats Inferray. In the issues query optimization and execution
Hiperfact's \emph{island processing} and \emph{derivation tree} components show a
concrete advantage over the competition. Future work for Hiperfact includes
dynamic caching of rank 2 and 3 query results, allowing fine grained result
among queries (including rule conditions), complex data types such as arrays
and dictionaries in the value part of a fact allowing for modeling of complex
data structures, type-based compression in the column-based join structures as
well as in the \emph{rank 1 index} to allow for more efficient storage and later
retrieval.

\bibliographystyle{splncs03}
\bibliography{refs}

\begin{thebibliography}{10}
\providecommand{\url}[1]{\texttt{#1}}
\providecommand{\urlprefix}{URL }

\bibitem{6291e0a3533703683d5355156812b697}
Abadi, D., Madden, S., Ferreira, M.: Integrating compression and execution in
  column-oriented database systems. In: Proceedings of the 2006 ACM SIGMOD
  International Conference on Management of Data. pp. 671--682. SIGMOD '06,
  ACM, New York, NY, USA (2006),
  \url{http://doi.acm.org/10.1145/1142473.1142548}

\bibitem{2cf321ecbafa720269187402d924b165}
Aref, M.M., Tayyib, M.A.: Lana-match algorithm: A parallel version of the
  rete-match algorithm. Parallel Comput.  24(5-6),  763--775 (Jun 1998),
  \url{http://dx.doi.org/10.1016/S0167-8191(98)00003-9}

\bibitem{efa74e5dfb852d2fbb9a48862a427843}
Balkesen, C., Alonso, G., Teubner, J., \"{O}zsu, M.T.: Multi-core, main-memory
  joins: Sort vs. hash revisited. Proc. VLDB Endow.  7(1),  85–96 (Sep 2013),
  \url{https://doi.org/10.14778/2732219.2732227}

\bibitem{b913d50abea5e224196416290c4b1e5e}
Blanas, S., Li, Y., Patel, J.: Design and evaluation of main memory hash join
  algorithms for multi-core cpus. pp. 37--48 (01 2011)

\bibitem{a7c876b59de9c3021c33a8b9c3350ff0}
Bramas, B.: {Fast Sorting Algorithms using AVX-512 on Intel Knights Landing}
  (Apr 2017), \url{https://hal.inria.fr/hal-01512970}, working paper or
  preprint

\bibitem{7a2b382cfad80e438d06946e27075f8d}
Brant, D.A., Grose, T., Lofaso, B., Miranker, D.P.: Effects of database size on
  rule system performance: Five case studies. In: Proceedings of the 17th
  International Conference on Very Large Data Bases. p. 287–296. VLDB ’91,
  Morgan Kaufmann Publishers Inc., San Francisco, CA, USA (1991)

\bibitem{20201204091851}
Drepper, U.: What every programmer should know about memory (2007)

\bibitem{36a96c21cb9e42c4bb5ffa0ddd53c243}
Erling, O., Averbuch, A., Larriba-Pey, J., Chafi, H., Gubichev, A., Prat, A.,
  Pham, M.D., Boncz, P.: The ldbc social network benchmark: Interactive
  workload. In: Proceedings of the 2015 ACM SIGMOD International Conference on
  Management of Data. p. 619–630. SIGMOD ’15, Association for Computing
  Machinery, New York, NY, USA (2015),
  \url{https://doi.org/10.1145/2723372.2742786}

\bibitem{8ac1b9e0e0ebe78e1c9f23ba31dadaf6}
Fabret, F., R\'{e}gnier, M., Simon, E.: An adaptive algorithm for incremental
  evaluation of production rules in databases. In: Proceedings of the 19th
  International Conference on Very Large Data Bases. p. 455–466. VLDB ’93,
  Morgan Kaufmann Publishers Inc., San Francisco, CA, USA (1993)

\bibitem{f6c08841b199d4e8326ce30f73705bf8}
Faye, D., Curé, O., Blin, G.: A survey of rdf storage approaches. ARIMA
  Journal  15,  11--35 (01 2012)

\bibitem{original_rete}
Forgy, C.: Rete: {A} fast algorithm for the many patterns/many objects match
  problem. Artif. Intell.  19(1),  17--37 (1982)

\bibitem{b260dee9ff2be53e2e7d030c020d45f9}
Gubichev, A., Neumann, T.: Exploiting the query structure for efficient join
  ordering in sparql queries. In: IN EDBT. pp. 439--450 (2014)

\bibitem{lubm}
Guo, Y., Pan, Z., Heflin, J.: Lubm: A benchmark for owl knowledge base systems.
  Journal of Web Semantics  3(2),  158 -- 182 (2005),
  \url{http://www.sciencedirect.com/science/article/pii/S1570826805000132},
  selcted Papers from the International Semantic Web Conference, 2004

\bibitem{91fd3608b3799a815471ec5708af5f6f}
He, L., Shao, B., Li, Y., Xia, H., Xiao, Y., Chen, E., Chen, L.J.: Stylus: A
  strongly-typed store for serving massive rdf data. Proc. VLDB Endow.  11(2),
  203--216 (Oct 2017),
  \url{https://doi-org.uaccess.univie.ac.at/10.14778/3149193.3149200}

\bibitem{ef4794de8d86f3e0882ea6b473e8d353}
Hill, E.F.: Jess in Action: Java Rule-Based Systems. Manning Publications Co.,
  USA (2003)

\bibitem{1c07684d12a395a56ede6eb1d63432a7}
Inoue, H., Taura, K.: Simd- and cache-friendly algorithm for sorting an array
  of structures. Proc. VLDB Endow.  8(11),  1274–1285 (Jul 2015),
  \url{https://doi.org/10.14778/2809974.2809988}

\bibitem{16f19f9427e3d380b0b164833731acfb}
Jin, C., Carbonell, J., Hayes, P.: Argus: Rete + dbms = efficient persistent
  profile matching on large-volume data streams. vol. 3488, pp. 142--151 (05
  2005)

\bibitem{9908f7b1914bdda05e147179e7d2c7c2}
Khadilkar, V., Kantarcioglu, M., Thuraisingham, B., Castagna, P.: Jena-hbase: A
  distributed, scalable and effcient rdf triple store. (01 2012)

\bibitem{3b364c343fbb1594558df47b2d56b5c3}
Kim, M., Lee, K.S., Kim, Y., Kim, T., Lee, Y., Cho, S., Lee, C.G.: Rete-adh: An
  improvement to rete for composite context-aware service. International
  Journal of Distributed Sensor Networks  2014,  1--11 (04 2014)

\bibitem{baeb5a24d59d88ccd391abfdcf2ae162}
Lang, H., Leis, V., Albutiu, M.C., Neumann, T., Kemper, A.: Massively Parallel
  NUMA-Aware Hash Joins, pp. 3--14 (01 2015)

\bibitem{dc7be68d49267fd0c3af2b1941810f87}
Liang, S., Fodor, P., Wan, H., Kifer, M.: Openrulebench: An analysis of the
  performance of rule engines. In: Proceedings of the 18th International
  Conference on World Wide Web. p. 601–610. WWW ’09, Association for
  Computing Machinery, New York, NY, USA (2009),
  \url{https://doi.org/10.1145/1526709.1526790}

\bibitem{openrulebench}
Liang, S., Fodor, P., Wan, H., Kifer, M.: Openrulebench: An analysis of the
  performance of rule engines. In: Proceedings of the 18th International
  Conference on World Wide Web. p. 601–610. WWW ’09, Association for
  Computing Machinery, New York, NY, USA (2009),
  \url{https://doi.org/10.1145/1526709.1526790}

\bibitem{456a57403b6029b0f1c5469bac921edd}
Marr, S., Renaux, T., Hoste, L., Meuter, W.: Parallel gesture recognition with
  soft real-time guarantees. Science of Computer Programming  98 (04 2014)

\bibitem{wordnet}
Miller, G.A.: Wordnet: A lexical database for english. Commun. ACM  38(11),
  39–41 (Nov 1995), \url{https://doi.org/10.1145/219717.219748}

\bibitem{rdfox}
Motik, B., Nenov, Y., Piro, R., Horrocks, I., Olteanu, D.: Parallel
  materialisation of datalog programs in centralised, main-memory rdf systems.
  In: Proceedings of the Twenty-Eighth AAAI Conference on Artificial
  Intelligence. p. 129–137. AAAI’14, AAAI Press (2014)

\bibitem{20201118205457}
Nam, Y.M.N., Han, D.H., Kim, M.S.K.: Sprinter: A fast n-ary join query
  processing method for complex olap queries. In: Proceedings of the 2020 ACM
  SIGMOD International Conference on Management of Data. p. 2055–2070. SIGMOD
  '20, Association for Computing Machinery, New York, NY, USA (2020),
  \url{https://doi.org/10.1145/3318464.3380565}

\bibitem{7548985}
{Nyman}, L., {Laakso}, M.: Notes on the history of fork and join. IEEE Annals
  of the History of Computing  38(3),  84--87 (2016)

\bibitem{06dde6ee870827728c0e31572574d9bb}
Proctor, M., Fusco, M., Sottara, D., Zimányi, T.: Reducing the Cost of the
  Linear Growth Effect Using Adaptive Rules with Unlinking and Lazy Rule
  Evaluation: Confederated International Conferences: CoopIS, C\&TC, and ODBASE
  2018, Valletta, Malta, October 22-26, 2018, Proceedings, Part II, pp.
  592--601 (10 2018)

\bibitem{7b4bf441c107aa7406e0480be13794de}
Riley, G.: Clips: An expert system building tool  (1991)

\bibitem{c5438a4cbac7b20c278ce34dd3ee1100}
Schmidt, K.U., St{\"u}hmer, R., Stojanovic, L., Anicic, D., Brelage, C.,
  Etzion, O., Stojanovic, N.: Blending complex event processing with the rete
  algorithm (2008)

\bibitem{4440df1a0c45c9dd3d8ad7c190c6d563}
{Sottara}, D., {Mello}, P., {Proctor}, M.: A configurable rete-oo engine for
  reasoning with different types of imperfect information. IEEE Transactions on
  Knowledge and Data Engineering  22(11),  1535--1548 (2010)

\bibitem{8f2df16f0f940d37bd677bd56e5ae115}
Subercaze, J., Gravier, C., Chevalier, J., Laforest, F.: Inferray: Fast
  in-memory rdf inference. Proc. VLDB Endow.  9(6),  468--479 (Jan 2016),
  \url{http://dx.doi.org/10.14778/2904121.2904123}

\bibitem{d521de21f7e948408283be47ca80e55b}
Zhou, R., Wang, G., Wang, J., Li, J.: Runes ii: A distributed rule engine based
  on rete network in cloud computing. International Journal of Grid and
  Distributed Computing  7,  91--110 (12 2014)

\end{thebibliography}

\end{document}